\documentstyle[psfig]{l-aa}

\begin{document}
\thesaurus{3(11.01.2; 11.09.1 WNB\,0313+683; 11.09.3; 13.18.1)} 
\title{WNB\,0313+683: Analysis of a Newly Discovered Giant Radio Galaxy}
\author{A.P. Schoenmakers\inst{1,5}
\and K.-H. Mack\inst{2,3}
\and L. Lara\inst{4}
\and H.J.A. R{\"o}ttgering\inst{5}
\and A.G. de Bruyn\inst{6,7}
\and H. van der Laan\inst{1}
\and G. Giovannini\inst{2,8}
}
\institute{Astronomical Institute, Utrecht University, P.O. Box 80\,000, 3508~TA Utrecht, The Netherlands
\and {Istituto di Radioastronomia del CNR, Via P. Gobetti 101, I-40129 Bologna, Italy}
\and {Radioastronomisches Institut der Universit\"at Bonn, Auf dem H\"ugel 71, D-53121 Bonn, Germany}
\and {Instituto de Astrofisica de Andalucia (CSIC), Apdo. 3004, 18080 Granada,
Spain}
\and {Sterrewacht Leiden, Leiden University, P.O. Box 9513, 2300~RA Leiden, The Netherlands}
\and {NFRA, P.O. Box 2, 7990~AA Dwingeloo, The Netherlands}
\and {Kapteyn Institute, University of Groningen, P.O. Box 800, 9700~AV Groningen, The Netherlands}
\and {Dipartimento di Fisica - Universit\`a di Bologna, Via B. Pichat 6/2, I-40127 Bologna, Italy}
}
\offprints{A.P. Schoenmakers}              
\date{Received <date> ; accepted <date>}
\maketitle

\begin{abstract}
 
We present the results of a detailed analysis of the newly discovered 15$'$ large radio galaxy WNB\,0313+683. It has been discovered in the WENSS and NVSS radio surveys, and has been identified with an optical galaxy at a redshift of $0.0901\pm0.0002$. The linear size of the radio structure is 2.0~Mpc, which places it in the class of Giant Radio Galaxies (GRGs).\\
Radio observations have been carried out with the WSRT at 1.4~GHz and with the 92-cm broadband system, with the 100-m Effelsberg telescope at 10.45~GHz and with the VLA at 1.4~GHz and 5~GHz.
At 10.45~GHz the core is the dominant source structure, emitting $\sim 25\%$ of the total flux. It has an inverted spectrum with a spectral index of $\alpha=+0.42 \pm 0.03$ (S$_{\nu}\propto \nu^{\alpha}$) between 1.4~GHz and 10.45~GHz, suggesting a very compact structure, although we can not rule out variability of the core luminosity.\\
The Rotation Measure distribution has been mapped using a new method and has been found to be very uniformly distributed over the source. It is therefore probably galactic in origin, with only a small contribution from the (surroundings of the) source itself.\\
From the spectral index distribution we have determined an upper limit on the spectral age of $1.4 \pm 0.1 \times 10^8$~yrs. The particle density of the ambient medium, using ram-pressure equilibrium at the hotspots, is $\ga 1.6 \times 10^{-6}$~cm$^{-3}$ for the southern lobe and $\ga 5.8 \times 10^{-7}$~cm$^{-3}$ for the northern lobe. An independent measure of the external density has been determined using the amount of depolarization towards the southern lobe. This gives a density, averaged along the line of sight, which is a factor of 10 higher ($2.0\times10^{-5}$~cm$^{-3}$) than the density near the head of the lobes found using the ram-pressure arguments. This discrepancy might at least partly be the result of a contribution from internal depolarization, which we can not exclude on basis of our radio data.\\
From spectroscopical observations of the host galaxy we find that the H$\alpha$ emission line has a broad component, and that the extinction must be large with colour index $E(B-V) = 0.98 \pm 0.10$~mag. Since the galactic latitude is $+9\fdg8$, the extinction is probably mostly galactic in origin.
We further find that WNB\,0313+683 has a very large optical emission line flux with respect to its estimated jet power, when compared with the correlation between these two properties found by Rawlings \& Saunders (1991). We argue that this, together with the relatively high radio power and the inverted radio spectrum of the radio core, is suggestive of a new phase of radio activity in WNB\,0313+683.\\

\keywords{Galaxies: active; Galaxies: Individual: WNB\,0313+683; Radio continuum: galaxies; Intergalactic medium}

\end{abstract}

\section{Introduction}

Giant Radio Galaxies (GRGs, e.g. Saripalli et al. 1986, Subrahmanyan et al. 1996) are the largest members of the radio galaxy population, with  a (projected) linear size $\ga 1.0$~Mpc\footnote{We use ${\rm H}_0 = 50$~km~s$^{-1}$~Mpc$^{-1}$ and ${\rm q}_0 = 0.5$ throughout this paper.}. At low redshift ($z \la 0.3$) some thirty of these large sources are known. The most extreme case is the radio source 3C\,236 at a redshift of $\sim 0.09$, which has a projected linear size of 5.7 Mpc (Willis et al. 1974, Strom \& Willis 1980, Barthel et al. 1984).\\
GRGs are interesting objects to study. First, at low redshift their angular size is several arcminutes or larger, which allows detailed studies of the different components of their radio structure, such as their jets (e.g. NGC\,6251, Perley et al. 1984) and their lobe emission (e.g. Mack et al. 1997a) using a variety of radio instruments. 
Secondly, because of the large physical size of GRGs, they have expanded well out of the denser central regions of the clusters that they reside in, into a much less dense Intergalactic Medium (IGM). This makes GRGs a powerful and unique tool to probe the low-density medium at a large distance from their host
galaxies, which is inaccessible to current X-ray instruments.
Thirdly, their active galactic nucleus (AGN) must be approaching the endstage of its radio active phase. GRGs can therefore provide important information on the properties of old AGN.  
Lastly, several of these large sources have been identified with quasars (e.g. 4C\,34.47, J{\"a}gers et al. 1982) or Broad Line Radio Galaxies (e.g. 0319+412, de Bruyn 1989). This makes them important test cases for the orientation dependent unification schemes for radio loud AGN (e.g. Barthel 1989). These state that quasars and radio galaxies are mostly alike, but that quasars have their radio jets oriented closer to the line of sight than radio galaxies. A natural consequence of this is that quasars are
 not expected to have large projected linear sizes.\\
There are now many indications that the ambient medium of the radio lobes of GRGs has a very low density ($\sim\!10^{-6} - 10^{-5}$~cm$^{-3}$).
Probably the strongest evidence is provided by observations of the Rotation Measure (RM) and the depolarization towards these sources.
The measured RMs are small, with values which are usually $\la 20$~rad~m$^{-2}$ (e.g. Klein et al. 1994, Strom \& Willis 1980). Depolarization occurs only at wavelengths above 10-20~cm (Willis \& Strom 1978, Strom \& Willis 1980, J{\"a}gers 1986, Klein et al. 1996). This combination can only be explained
satisfactorily by a very low density of thermal electrons along the line of sight.\\ 
Further indications for a low density environment result from spectral age analyses of GRGs. Typical spectral ages for GRGs have been found to be $10^7 - 10^8$~yrs (e.g. Mack et al. 1998), which translates into expansion velocities of $0.01c - 0.1c$. Using the assumption of ram pressure equilibrium at the head of the jet (e.g. Miley 1980), external densities of $\sim10^{-6} - 10^{-5}$~cm$^{-3}$ are commonly found for GRGs. This method suffers from many assumptions that have to be made, such as equipartition between the energy density of the relativistic particles
and the magnetic field, the filling factor of the radio lobes, the fraction of
energy in heavy particles, the area of the bow shock, and so on. It is therefore often criticized (e.g. Eilek et al. 1997) and its result should be interpreted with care.\\
Many clusters have large and bright X-ray haloes, often extending to distances of more than 1~Mpc from their centers and containing cooling flows (e.g. Fabian 1994). X-ray studies of some GRGs (e.g. NGC~6251; Mack et al. 1997b), however, have found only weak thermal X-ray emission around the host galaxies. GRGs are therefore probably not inside rich clusters with dense cores, and the lobes are thus not likely to be found in dense environments.
Subrahmanyan et al. (1996) studied a small sample of GRGs on the southern hemisphere. By studying the surface density of optical galaxies in the neighbourhood of the GRG host galaxies using the UKST plates, they conclude that they do not reside in rich clusters.\\
GRGs have relatively low radio powers (Saripalli et al. 1986; Subrahmanyan et al. 1996), usually around or below the luminosity which divides FRI and FRII type radio galaxies ($\sim 2 \times 10^{26}$~W~Hz$^{-1}$ at 178~MHz; Fanaroff \& Riley 1974). Because of their large sizes and relatively low radio power, the surface brightness of GRGs is low. This is why they are so difficult to find in most radio surveys.
In recent years, the Westerbork Northern Sky Survey (WENSS, Rengelink et al. 1997) has mapped the sky above +28\degr~declination at a frequency of 327~MHz. WENSS has a sensitivity of 18~mJy (5$\sigma$) and a beamsize of $54\arcsec \times 54\arcsec~{\rm cosec}\,\delta$, with $\delta$ the declination. At low frequency, the bulk of emission of radio galaxies originates in the extended radio lobes. This makes the WENSS ideally suited to find
large, low surface brightness radio galaxies such as GRGs. The first
discovery of such an object (WNB\,1626+5153) has been reported by R\"{o}ttgering et al. (1996). This initiated a project aimed at finding and studying a large, uniformly selected, sample of giant radio sources from the WENSS.\\
Here we report on the discovery and subsequent analysis of the radio source WNB\,0313+683, which we have identified as an FRII-type radio galaxy with a linear size of 2.0~Mpc. It has several remarkable properties, among which there is a large flux asymmetry of the radio lobes and a prominent, inverted spectrum, radio core. In Sect. 2, we will present the radio and optical data that we have collected on this object. Section 3 describes a first analysis of these data, including a new way to measure Rotation Measures. It also presents some of the derived physical
properties. In Sect. 4 we derive the advance velocities of the hotspots and the age of the source from a spectral index analysis. Section 5 then discusses the observed depolarization towards WNB\,0313+683. A discussion on the properties of the radio core is given in Sect. 6. We argue that WNB\,0313+683 may currently be in a new phase of radio activity. Finally, a summary and our conclusions are presented in Sect. 7.

\section{Observations}

\begin{table*}
\caption{\label{tab:radio_obs} }
\begin{flushleft}
\begin{scriptsize}
\begin{tabular}{l l l l l l}
\hline \hline \\
\multicolumn{6}{c}{Non-survey radio observations of WNB\,0313+683}\\
\hline \\
\multicolumn{1}{c}{Telescope} & \multicolumn{1}{c}{Date} & \multicolumn{1}{c}{Freq.} & \multicolumn{1}{c}{Bandwidth} & \multicolumn{1}{c}{Int. time} & \multicolumn{1}{c}{r.m.s.}\\
\hline \\
WSRT & Aug. 6 1995 & 1435~MHz & $5\times 10$~MHz & 32 mins. ($8\times 4$) & 0.3 mJy \\
WSRT & Oct. 27, Nov. 11 1995 & 343~MHz & $8\times 5$~MHz & 9 hrs. ($6+3$) & 0.5, 1.4, 0.7 mJy (Stokes $V,Q,U$)\\
Effelsberg 100-m & Dec. 10-27 1995 & 10.45~GHz & 300~MHz & 1 min beam$^{-1}$ & 1.3 mJy (Stokes $I$), 0.37 mJy (Stokes $Q,U$)\\
VLA B-conf. & Nov. 19 1995 & 1425~MHz & $2\times 25$~MHz & 7 mins. & Combined with C-conf. data\\
VLA C-conf. & Feb. 19 1996 & 1425~MHz & $2\times 25$~MHz & 10 mins. & 0.13~mJy (NA-weighting)\\
VLA C-conf. & Feb. 19 1996 & 4860 MHz & $2\times 50$~MHz & 10 mins. & 0.04 mJy (NA-weighting)\\
\ \\
\hline \hline\\
\end{tabular}
\end{scriptsize}
\end{flushleft}
\end{table*} 

In this section we describe the observational data we have obtained for the source WNB\,0313+683. An overview of radio observations which are not related to publicly available surveys (i.e. the WENSS and the NVSS) can be found in Tab. \ref{tab:radio_obs}.   
 
\subsection{Survey data: WENSS and NVSS}

\begin{figure*}
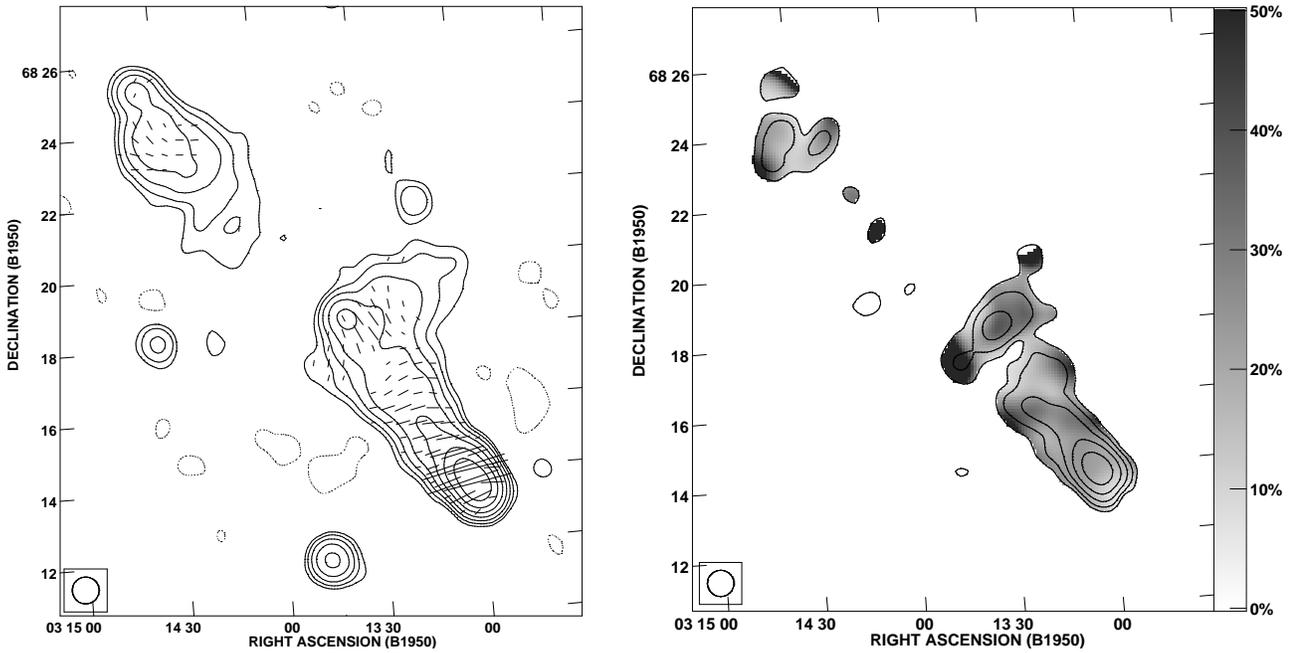

\begin{tabular}{ll}
\psfig{figure=7509.1a,width=0.437\textwidth,clip=} & \psfig{figure=7509.1b,width=0.5\textwidth,clip=} \\
\end{tabular} 
\caption{\label{fig:nvss}
Contour plots of the radio source WNB\,0313+683 at 1400 MHz, from the NVSS survey. {\bf a} Contours of total intensity and the $E$-vectors superposed. Contours are at -1.5,1.5,3,6,12,24,48,96,192 mJy~beam$^{-1}$. The length of the vectors represents the polarized intensity ($1\arcsec = 0.2$ mJy~beam$^{-1}$). {\bf b} Contours of polarized intensity with the fractional polarization in greyscale. The greyscale ranges from 0\% (white) to 50\% (black). Contourlevels are at 1.2,2.4,4.8,9.6,19.2 mJy~beam$^{-1}$.}
\end{figure*}

The radio source WNB\,0313+683 was noticed as a 15\arcmin~large FRII-type radio galaxy in the WENSS and in the NRAO VLA Sky Survey (NVSS; Condon et al. 1993). 
Figure \ref{fig:spec_indexmaps}a shows the radio contours from a map of WNB\,0313+683 from the WENSS survey. Clearly visible are the two bright regions at the outer edges of the radio lobes, the bridge connecting these regions, and a bright central lateral component extending to the north-west.  
The brightness and narrowness of the southern lobe suggests the presence of a radio jet. The northern lobe shows a protrusion at the top. From the WENSS map alone it is not clear if this protrusion is part of the radio galaxy (i.e. a hotspot), or due to an unrelated radio source. The total integrated flux density at 327~MHz is $3.12 \pm 0.15$~Jy. Flux densities for the different source components are given in Tab. \ref{tab:fluxes}.\\
The angular resolution of the NVSS survey is 45\arcsec, comparable to that of the WENSS survey. The typical noise in the produced maps is 0.45~mJy~beam$^{-1}$. Further, the NVSS provides maps of the Stokes parameters $Q$ and $U$.
Figure \ref{fig:nvss} shows the radio map from the NVSS survey, with vectors of the observed $E$-field superposed.   
The southern lobe and the central `bulge' are still well detected. The bridge to the northern lobe is very faint, which is a first indication of its steep spectrum. The polarization angles in the southern lobe are more or less constant in orientation; only near the core and the bulge there is a gradual change in polarization angle.\\
The weak sources at RA~$03^{\rm h}14^{\rm m}34\fs5$, Dec.~$68\degr18\arcmin10\arcsec$ and RA~$03^{\rm h}13^{\rm m}13\fs6$, Dec. $68\degr21\arcmin37\arcsec$ (B1950.0) are most probably unrelated background sources.\\ 

\subsection{WSRT Observations}

\subsubsection{WSRT 1.4 GHz snapshot observations}

\begin{figure*}
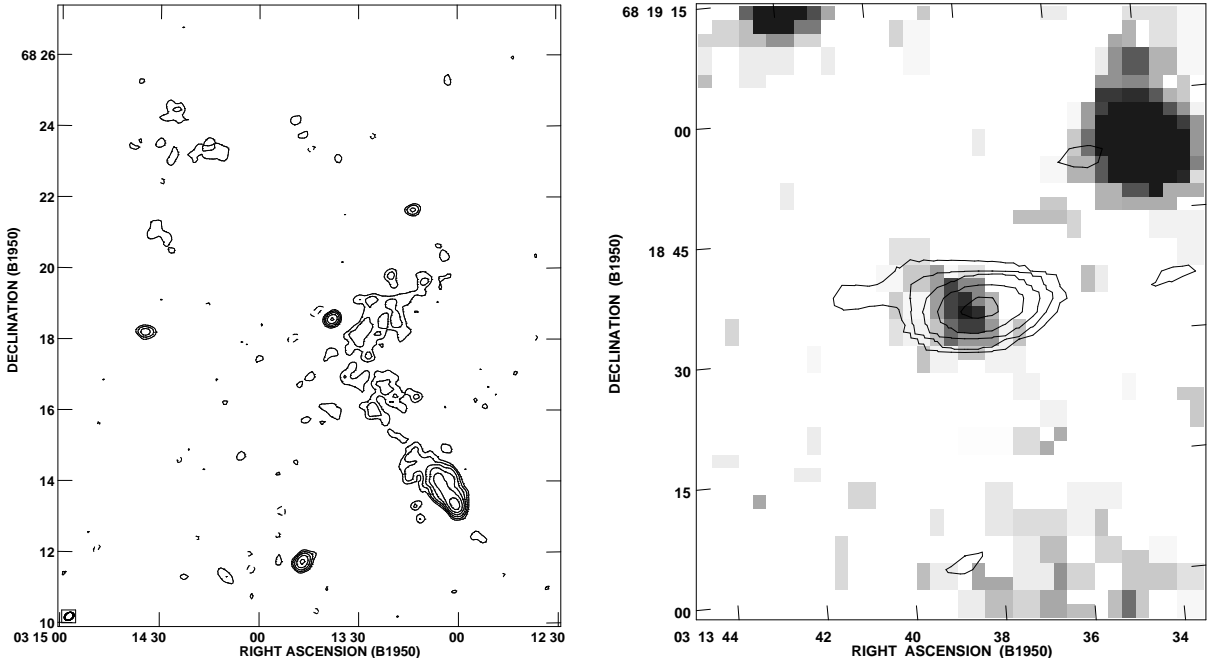

\begin{tabular}{ll}
\psfig{figure=7509.2a,width=0.421\textwidth,clip=} & 
\psfig{figure=7509.2b,width=0.45\textwidth,clip=}
\end{tabular}
\caption{\label{fig:wsrt1.4+opt}{\bf a} Contour map of the 1.4~GHz WSRT snapshot observations. Contour levels are at -1,1,2,4,8,16,32,64 mJy beam$^{-1}$. {\bf b} Overlay of the 5-GHz VLA radio map of the core of WNB\,0313+683 (contours) and an optical image extracted from the Digitized Sky Survey (greyscale). The radio core is identified with a faint and extended galaxy-like object.}
\end{figure*}

To find an accurate position of the radio core to facilitate identifying the host galaxy, we have observed WNB\,0313+683 with the Westerbork Synthesis Radio Telescope (WSRT) at 1.4~GHz.
We obtained a short observation on August 6, 1995, consisting of $8 \times 4$~minutes of integration, separated by typically 1.5 hours. Only a short (2~min) observation was made of the primary calibrator 3C\,286, which did not allow accurate ($\la 10\%$) flux calibration. Furthermore, the sparse UV-coverage (basically 8 spikes at regular hour-angle intervals) is not adequate to obtain a good radio map of all the sources structures. Therefore, we do not quote any flux values from these observations.\\
Calibration and mapping of the data were done using the NFRA {\sc newstar} package. 
The resulting map is presented in Fig. \ref{fig:wsrt1.4+opt}. A radio core is clearly detected and it is unresolved. Its B1950.0 position is $03^{\rm h}13^{\rm m}37\fs90\pm 0\fs05$ in right ascension and $+68\degr18\arcmin34\farcs08\pm 0\farcs2$ in declination. The southern hotspot region is resolved and elongated along the radio axis. A radio jet has not been detected in these observations.\\ 

\subsubsection{WSRT 92-cm broadband observations}

\begin{figure*}
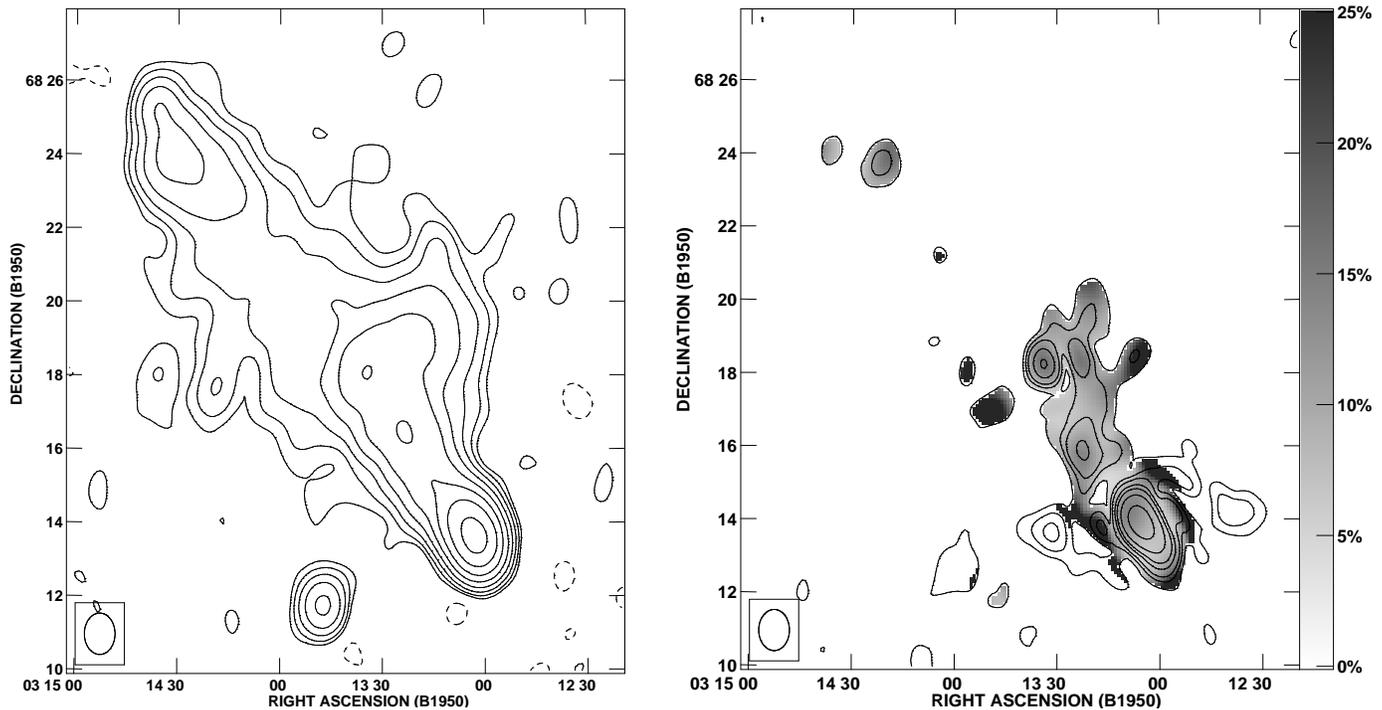

\begin{tabular}{ll}
\psfig{figure=7509.3a,width=0.4675\textwidth,clip=} & 
\psfig{figure=7509.3b,width=0.535\textwidth,clip=}
\end{tabular}
\caption{\label{fig:wsrt_92cm}{\bf a} Contour map of the 92-cm broadband observations. The map shown here is the average of the 5 good channels in these observations, and represents the source at 343~MHz. Contour levels are at -4,4,8,16,32,64,128,256,512 mJy beam$^{-1}$. {\bf b} Contour map of the polarized intensity, averaged over the 5 good channels of the 92-cm broadband observations. Contour levels are at 5,7.5,10,15,20,30,50,100 mJy beam$^{-1}$. The greyscale represents the fractional polarization, on a scale from 0\% (white) to 25\% (black).}
\end{figure*}

We observed WNB\,0313+683 using the 92-cm broadband system of the WSRT, to map the extended regions with low surface brightness in more detail. The 92-cm
broadband system uses 8 channels of 5 MHz, positioned between 319~MHz (94.0~cm wavelength) and 380~MHz (78.9~cm wavelength). 
We observed WNB\,0313+683 for 9 hours, distributed over two periods of 12 hours. The primary calibrators 3C\,48 and 3C\,286 were observed, which we used for gain and phase calibration. We use the scale of Baars et al. (1977). 
Due to man-made radio interference, only 5 of the 8 channels (the ones between 325~MHz~ and 360~MHz) could be used for mapping and further analysis of the data. Only one step of phase selfcalibration was necessary to improve the data quality to nearly the theoretical limit. Due to possible confusion by weak background sources in the total intensity map, the best estimate of the noise is obtained from the Stokes $V$ map. We measured a r.m.s. noise of $\sim 1$~mJy in each channel, and $\sim 0.5$~mJy in the average of the five useful channels.\\ 
For each channel, maps were made in Stokes $I$, $Q$, and $U$ parameters. The total power maps have been averaged into one single map, which is represented in Fig. \ref{fig:wsrt_92cm}. The sensitivity of the new observations is 6 times better than that of the WENSS survey. We detect additional emission outside of the lowest WENSS contours.\\
Even at a wavelength of 92 cm there is still a substantial amount of polarization measured.
We have mapped the polarized intensity in each individual channel, and averaged the results from the five good channels in one single map. The result is also shown in Fig. \ref{fig:wsrt_92cm}. Around the southern hotspot there are some residual sidelobes visible in the polarized intensity map. We decided not to
remove them, because they fall mainly outside the total intensity contours and are therefore easily discernible. We have not plotted the polarization vectors, because the amount of rotation that has occured at such a low frequency deprives this of any physical meaning. 

\subsection{Effelsberg observations}

\begin{figure*}
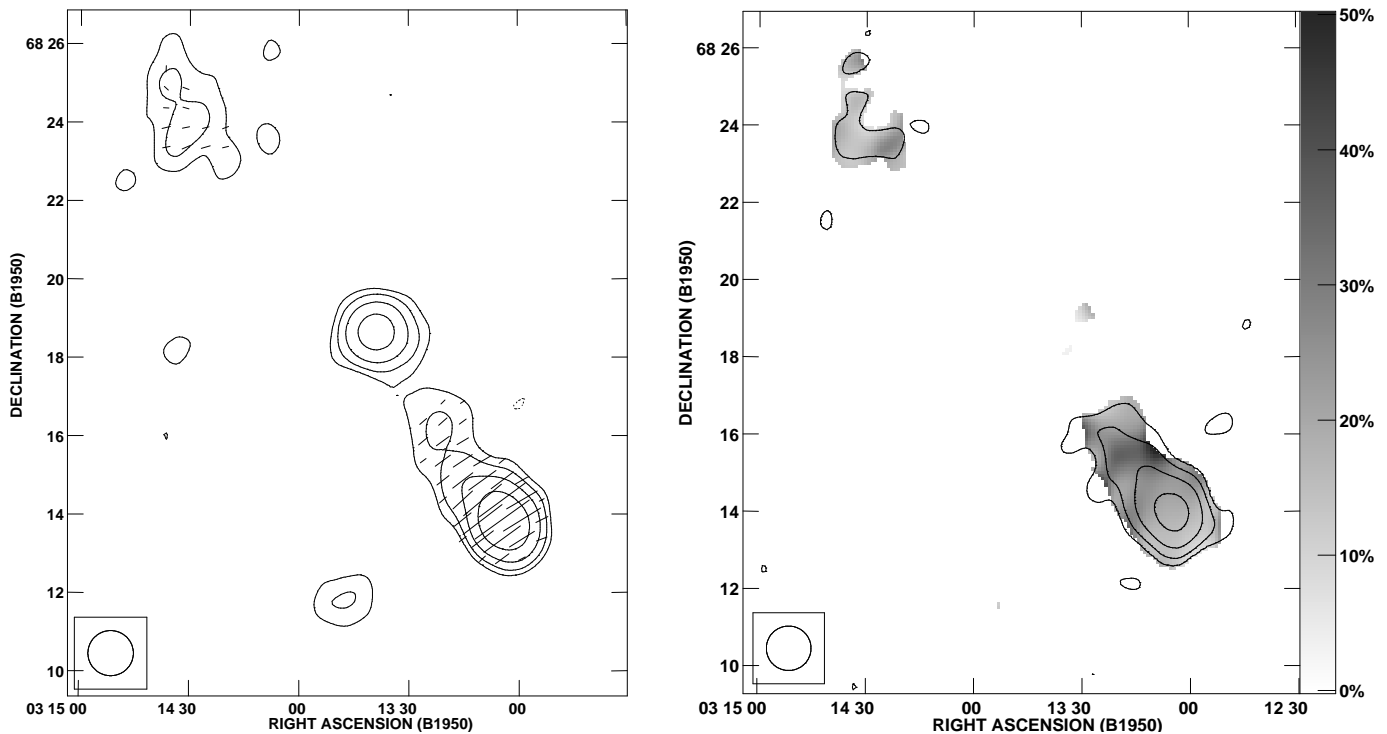

\begin{tabular}{ll}
\psfig{figure=7509.4a,width=0.4693\textwidth,clip=} & \psfig{figure=7509.4b,width=0.535\textwidth,clip=}\\
\end{tabular}
\caption{\label{fig:effelsberg} {\bf a} Contour map of the 10.45-GHz Effelsberg observations of WNB\,0313+683, with the measured $E$-vectors superposed. Contours are at -4,4,8,16,32 mJy beam$^{-1}$. The length of the vectors represents the polarized intensity ($1\arcsec = 0.1$~mJy beam$^{-1}$). {\bf b} Contour map of the polarized intensity at 10.45~GHz. Contours are at 1,2,4,8~mJy beam$^{-1}$. The greyscale represents the fractional polarization and ranges from 0\% (white) to 50\% (black).}
\end{figure*}

To study the high-frequency morphology and polarization properties of WNB\,0313+683, it was observed at 10.45~GHz with the 100-m Effelsberg telescope in December 1996. At this frequency the beamsize is $69''$, which is close to the WENSS beam. Furthermore, since the observations are done with a single-dish telescope, they are sensitive to large-scale structures, in contrary to most interferometers at this frequency. 
The observational and data reduction procedures are detailed by Gregorini et al. (1992).  The data have been calibrated according to the scale of Baars et al. (1977). In addition, the maps have been {\sc clean}ed, applying the algorithm described by Klein \& Mack (1995). 
The final maps are shown in Fig. \ref{fig:effelsberg}. The noise is 1.3~mJy beam$^{-1}$ for total, and 0.37~mJy beam${-1}$ for polarized intensity.
The total intensity map is dominated by the core and the southern lobe. We measure a flux density of the core of $55 \pm 3$~mJy at 10.45~GHz.\\ 
Polarized emission is detected in the radio lobes, but not towards the core. The southern radio lobe shows polarization degrees of up to 40\%. At this high frequency, the angles of the observed $E$-vectors are very close to their true direction\footnote{Faraday rotation rotates the polarization angle over $\Delta\vartheta = RM\cdot\lambda^2$~rad. For $|RM| < 50$~rad~m$^{-2}, \Delta\vartheta < 2\degr$, which is below the accuracy of our observations.}, 
so that we do not have to correct for Faraday rotation to derive the direction of the projected magnetic field in this source. The observed $E$-vectors are perpendicular to the radio axis, which means that the projected magnetic field must be largely parallel to the radio axis. This is comparable to what is generally found in powerful FRII radio sources (e.g. Saikia \& Salter 1988). 

\subsection{VLA observations}

\begin{figure*}
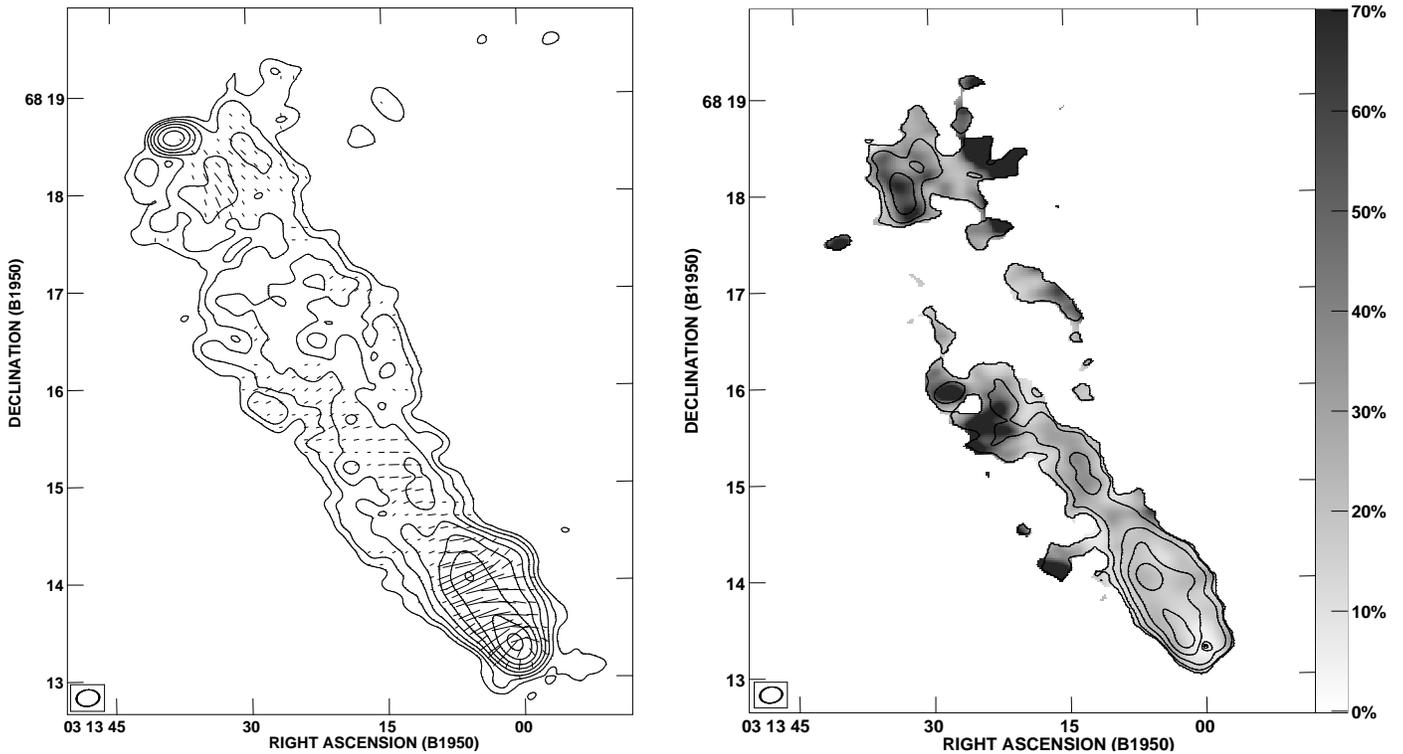

\begin{tabular}{ll}
\psfig{figure=7509.5a,width=0.475\textwidth,clip=} & \psfig{figure=7509.5b,width=0.54\textwidth,clip=} \\
\end{tabular}
\caption{\label{fig:VLA_pol_south}
{\bf a} Contour map of the southern radiolobe from our 1.4-GHz VLA observations. Contour levels are at -0.4,0.4,0.8,1.6,3.2,6.4,12.8,25.6,50.2 mJy~beam$^{-1}$. Superposed are the electric vectors of the polarized emission. The length of the vectors is proportional to the polarized intensity ($1\arcsec = 0.18$ mJy~beam$^{-1}$). {\bf b} Contour map of the polarized intensity. Contour levels are at 2.4,4.8,9.6,19.2,38.4 mJy~beam$^{-1}$. The greyscale represents the fractional polarization. The range is 0\% (white) to 70\% (black).}
\end{figure*}

\begin{figure*}
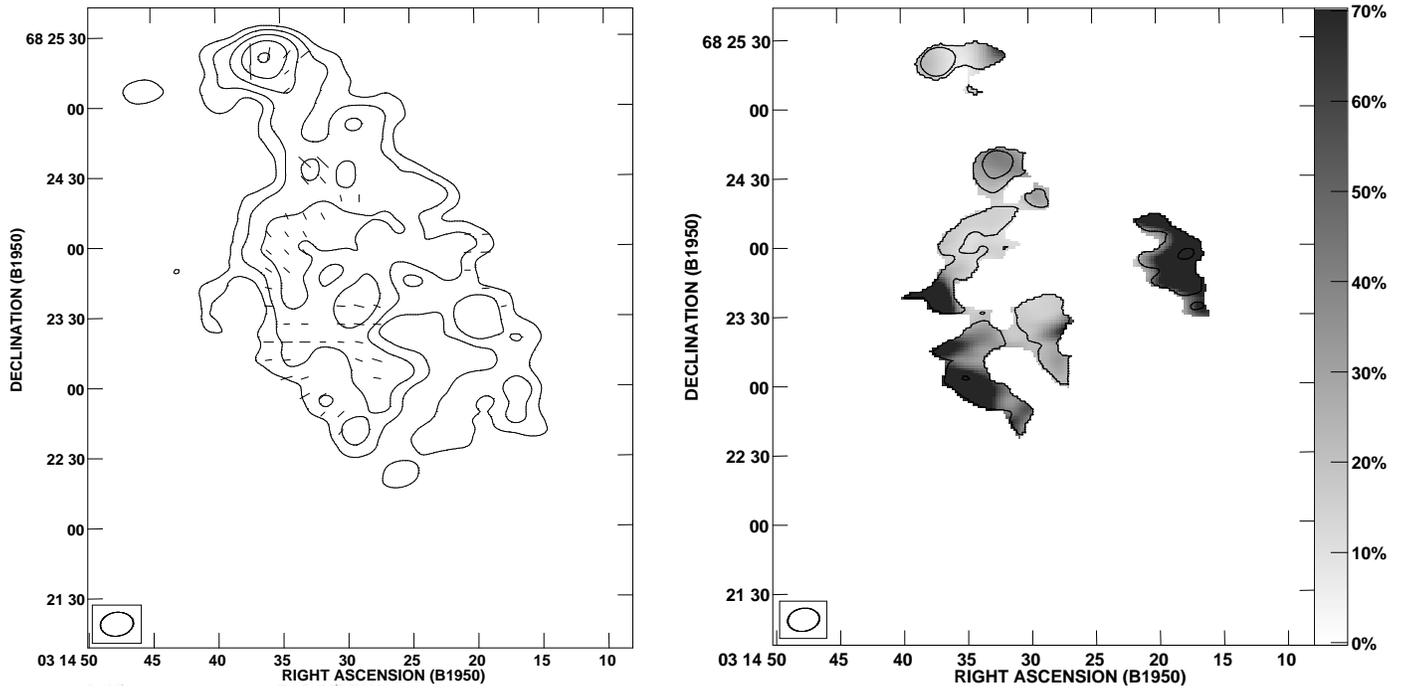

\begin{tabular}{l r}
\psfig{figure=7509.6a,width=0.4735\textwidth,clip=} & \psfig{figure=7509.6b,width=0.54\textwidth,clip=}\\
\end{tabular}
\caption{\label{fig:VLA_pol_north}
{\bf a} Contour plot of the northern radiolobe from our 1.4~GHz VLA observations. Contour levels are at -0.4,0.4,0.8,1.6,3.2,6.4 mJy~beam$^{-1}$. Superposed are the electric vectors of the polarized emission. The length of the vectors is proportional to the polarized intensity ($1\arcsec = 0.09$ mJy beam$^{-1}$). {\bf b} Contour plot of the polarized intensity. Contour levels are at 2.4,3.6,4.8,7.2 mJy~beam$^{-1}$. The greyscale represents the fractional polarization. The range is 0\% (white) to 70\% (black).}
\end{figure*}

The source WNB\,0313+683 was observed with the VLA in its B and C configuration on November 19th 1995 and February 19th 1996, respectively. 
We observed WNB\,0313+683 for 7 minutes in the B configuration, and for 10 minutes in the C configuration, using two standard 25-MHz IF bandwidths at frequencies of 1385 and 1465~MHz. Furthermore, we made a 10-min observation in the C configuration with two 50-MHz IFs at frequencies of 4835 and 4885 MHz.
The phases were calibrated observing  the nearby source 0217+738, and the radio sources 3C\,286 and 3C\,48 were used as primary flux density calibrators. Calibration and mapping of the data was carried out with the NRAO {\sc aips} package. 
To produce the final maps, the 1.4-GHz data from the B and C configurations were combined. 
The data have been mapped and selfcalibrated to solve for phase and gain variations during the observations. 
Total intensity and polarization maps have been produced, after {\sc clean}ing of the source structure using Natural Weighting of the UV-data (see Figs. \ref{fig:VLA_pol_south}, southern lobe, and \ref{fig:VLA_pol_north}, northern lobe).
At 1.4 GHz, both lobes are mapped in high detail. The southern lobe appears well confined; the width changes only by some 20\% along its length, and it has sharp outer edges. The northern lobe is more diffuse and its hotspot is much less bright.
The 5-GHz observations only reveal an unresolved core and no other source structures. The flux density of the core at this frequency is $40 \pm 3$~mJy.\\
Also shown in Figs. \ref{fig:VLA_pol_south} and \ref{fig:VLA_pol_north} are maps of the fractional polarization at 1.4 GHz. 
In the southern lobe, near the radio core, there is a region of strong fractional polarization with values reaching 60-70\%, which is close to the theoretical maximum. This must therefore be a region with a highly ordered magnetic field. The orientation of the observed $E$-field is here perpendicular to that in the southern lobe. Halfway the southern lobe there are two regions which also have a high fractional polarization. Although the outer edges of a radio lobe are often spuriously highly polarized, the extent of these two areas is too large to be ascribed to this effect, and we believe the increase in fractional polarization to be real. In the northern lobe only very little polarized structure is seen. The radio core is unpolarized at our detection limit. Its fractional polarization is therefore below 5\%, which is expected in the cores of radio galaxies (Saikia \& Salter 1988).\\

\begin{figure*}
\vbox{\psfig{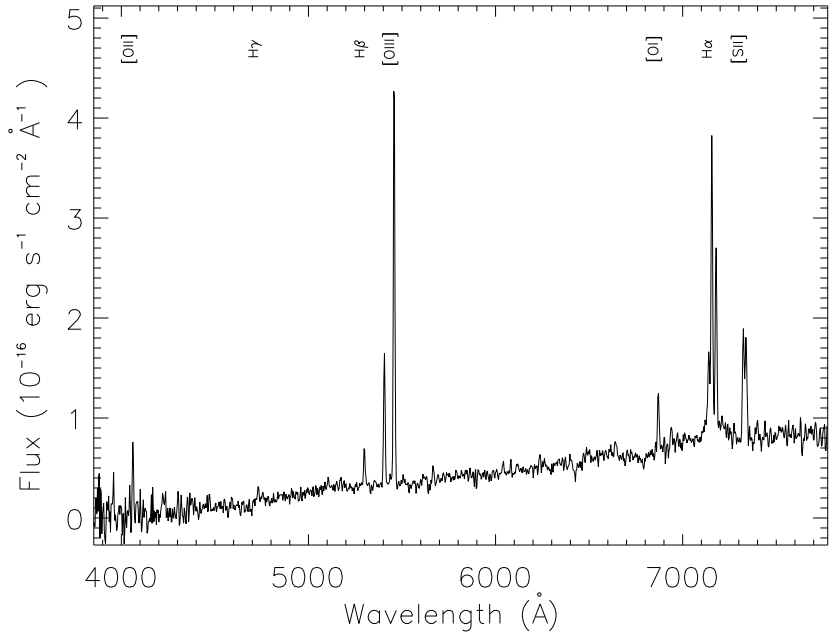}\vspace{-3.0cm}}
\hfill\parbox[b]{6.0cm}{\caption{\label{fig:ids-spectrum}
The optical spectrum of the host galaxy of WNB\,0313+683, corrected for atmospheric absorption features. Details on the observation can be found in the text. The most prominent emission lines have been indicated.}}
\end{figure*}

\subsection{Optical Observations}

The host galaxy was first identified by making an overlay of the WSRT 1.4-GHz map and the optical field obtained from the Digitized Sky Survey (the red POSS I plates). At the position of the radio core a weak optical galaxy is seen (see Fig. \ref{fig:wsrt1.4+opt}). Its B1950.0 coordinates are $03^{\rm h}13^{\rm m}38\fs33 \pm 0\fs08$ in RA and $68^{\circ}18'34\farcs3 \pm 1\farcs0$ in declination. The higher resolution 5-GHz VLA observations later confirmed the identification (see Fig. \ref{fig:wsrt1.4+opt}).\\  
We observed this galaxy with the 2.5-m INT telescope on La Palma, using the IDS spectrograph equipped with a 1k$\times$1k TEK chip. A pixel scale of 3.12\,\AA/pixel in  wavelength was used, resulting in a total wavelength coverage of $\sim\,3100$\AA. 
The spatial scale along the slit is $0\farcs84$/pixel. The orientation of the slit was parallel to the radio axis, and a slitwidth of 2\arcsec~was used. This gives a spectral resolution of $\sim 10$\AA.\\
One 1200-s exposure was made in August 1995, with a central wavelength of 5500\AA. In October 1996, two additional 600-s exposures were made at a central wavelength of 6000\AA. After the galaxy exposure, an exposure of a nearby F8-type star was made. This we used to correct for the atmospheric absorption bands.
All spectra were reduced in the standard way using the {\sc longslit} package in the NOAO {\sc iraf} data reduction software. 
Wavelength calibration was done using arc-lamp exposures and checked against the skylines on the original frames. Flux calibration was achieved by observations of several spectrophotometrical standard stars during the night. 
In the final extraction a $4''$ spatial aperture was used. This corresponds to 9~kpc at the redshift of the galaxy. 
The resulting spectrum is shown in Fig. \ref{fig:ids-spectrum}. It shows a wealth of emission lines, typical of a narrow-line active galactic nucleus (AGN), on top of a weak stellar continuum. Using the emsiion-lines we measure a redshift of $0.0901\pm0.0002$.

\section{Results}

In this section we determine several source characteristics from the available data. First, we discuss the overall radio morphology and some properties of the radio core. Secondly, we produce spectral index maps using the WENSS, NVSS and Effelsberg data. 
Thirdly, we investigate the radio polarization properties of the source. We apply a new method for finding Rotation Measures and we present maps of the observed depolarization towards WNB\,0313+683.
Then, we derive some general physical properties of the source from the radio data, such as internal energy densities and pressures. Lastly, we analyse the optical spectrum in more detail.

\subsection{Morphology, fluxes, and the radio core}

\begin{figure}
\psfig{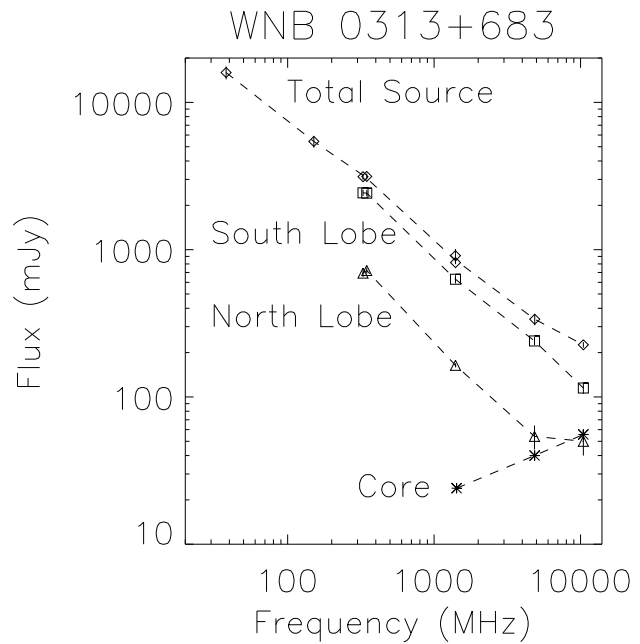}
\caption{\label{fig:fluxes}Radio continuum spectrum of WNB\,0313+683 for the whole source, the southern lobe, the northern lobe and the radio core. See Tab. 2 for the flux densities and their references.}
\end{figure}

The morphology, although typical of an FRII radio galaxy, shows some interesting features especially at 327 MHz. As mentioned, the most prominent feature is the bright narrow southern lobe, and the bright region at the end of it. The VLA observations have
 resolved this hotspot into two distinct bright compact components with a narrow spur of emission between them. It might be the case that WNB\,0313+683 has two hotspots in its southern lobe, similar to what is seen, for instance, in the well studied case of Cygnus A (e.g. Carilli et al. 1991). The difference is that in WNB\,0313+683 the two hotspots are well aligned with the radio axis.\\
The northern lobe can be roughly divided into two parts: the brighter northernmost region and the much weaker extended emission that fills up the area between the core and the brighter part (the bridge). Although there is a hotspot, it is not as powerful as its southern counterpart. The radio lobe as a whole has a  much more relaxed morphology.\\
An important parameter to describe the structural evolution of radio galaxies is the ratio between the length and the width of the radio source, the so-called axial ratio (e.g. Leahy \& Williams 1984). A low axial ratio indicates a rapid lateral expansion of the radio emitting plasma, most likely caused by a high internal pressure in the radio lobes, as compared to the pressure in the ambient medium (e.g. Begelmann \& Cioffi 1989; Nath 1995). The axial ratio can best be measured at a low frequency, because of the increased visibility of the old, extended regions.
Using the method described by Leahy \& Williams (1984), we find a value of $\sim 6.7$ from our 92-cm WSRT observations. This is similar to the axial ratios of other GRGs (Subrahmanyan et al. 1996) which are typically between 4 and 11.
It suggests that WNB\,0313+683 has a similar internal pressure as other GRGs. It will be shown later that the equipartition pressures in the radio lobes of WNB\,0313+683 are $\sim 10^{-14}$~dyn~cm$^{-2}$, which is indeed within the range found for other GRGs (Subrahmanyan \& Saripalli 1993, Mack et al. 1998).\\
Table \ref{tab:fluxes} gives the flux densities of the different components of WNB\,0313+683 at different frequencies, from our own data and from the literature and on-line databases. Where possible, we have made a distinction between the northern lobe, the southern lobe and the core. If no value for the core is given, it is included in the flux density of the southern lobe.\\ 
The NVSS flux density, although originating from interferometer data, agrees
well with the value from the single-dish NRAO 1400-MHz survey (Condon \& Broderick 1985). The beam of the latter is $\sim 13' \times 11'$ so that the three background sources visible in the NVSS map are incorporated in the source.
We have estimated the flux density of WNB\,0313+683 by subtracting the NVSS flux densities of the background sources from the NRAO 1400-MHz flux density. The NVSS flux density of WNB\,0313+683 is still somewhat lower than this estimate, which can be expected for a source of this angular extent.\\
Figure \ref{fig:fluxes} shows a plot of the values given in Tab. \ref{tab:fluxes}.
Note the upturn of the spectrum of the northern lobe at 10.4~GHz. A similar effect has been observed in the GRG 1358+305 (Parma et al. 1996), which was also observed with the Effelsberg telescope at the same frequency.\\
From Fig. \ref{fig:fluxes}, it appears that the radio core has an inverted spectrum.
But since the data have been taken at different epochs, it can also be variable. To investigate this we use the 1.4-GHz WSRT and VLA observations, which were done some $3-6$ months apart.
Although the flux density calibration of the short 1.4-GHz WSRT observations was not very accurate, we compared the WSRT fluxes of three background sources in the field and of the core with the VLA observations. In case of no variability, the ratio of the VLA and WSRT flux densities should be the same within the errors. This is indeed the case for the three background sources, but the core flux ratio differs from the mean of the background sources by $\sim 2\sigma$. To calculate the errors in these values we assumed a 2\% uncertainty in the flux density calibration of the VLA data and a 10\% uncertainty in the WSRT data.
Part of the deviation of the radio core flux ratio can be explained by the VLA observations picking up more of the extended emission surrounding the core. In the WSRT data, the UV-plane is poorly sampled and the map misses much of the extended structure. Therefore, we do not think that the found variability is
significant, although we cannot rule it out either. Additional data is needed to put more stringent constraints on this.\\
Neglecting the possible variability, we find a core spectral index of $+0.42 \pm 0.03$ between 1400 and 10450 MHz from a least-square fit through the three available flux density measurements. The turnover in the spectrum of the core must be at a frequency above 10 GHz.\\ 
We can estimate its size by assuming that the inverted spectrum is caused by synchrotron self-absorption.
The equipartition angular size $\vartheta_{eq}$ of the core is the size for which a source with given peak flux density and peak frequency is in equipartition (Scott \& Readhead 1977). Assuming that the spectral index above the turnover is $\sim -0.75$, we find $\vartheta_{eq} \sim 10^{-4}\arcsec$. This result varies very little on the chosen spectral index.
At the redshift of this source, $10^{-4}\arcsec$ translates into a linear size of only $\sim 0.2$~pc. \\ 
The fraction of the total emission that comes from the core, or the core-ratio $C$ (e.g. Orr \& Browne 1982), is a strongly increasing function of frequency for WNB\,0313+683: At 1400~MHz, $C = 0.03$, at 4850~MHz, $C = 0.12$ and at 10450~MHz, $C = 0.25$.  
The median value for 3CR radio galaxies is only $0.002$ at 8~GHz (Saikia \& Kulkarni 1994). Values as high as $\sim 0.2$ at 8~GHz are normally only seen in 3CR quasars. 
Saripalli et al. (1997) find that in GRGs a typical value for $C$ at 1.7~GHz is $C \sim 0.01$, and probably even somewhat larger at higher frequencies due to the usually steep spectrum of the radio lobes. Still, a core-ratio of $\sim 0.2$ at 8~GHz is very high, even for GRGs.\\
The radio core prominence can be used as an orientation indicator in the light
of relativistic beaming and unified scheme models (e.g. Giovannini et al. 1994,
Morganti et al. 1997). Comparing the measured core radio power with the
expected intrinsic core radio luminosity from the general correlation between
the core and the total radio power (Giovannini et al. 1988) we can derive
constraints on the orientation with respect to the plane of the sky of the arcsecond core emission. Objects with a core radio power stronger than the expected value are interpreted as galaxies where the core radio emission is Doppler-boosted by a relativistic parsec scale jet (see Giovannini et al. 1994 for a more detailed discussion). Comparing the measured core emission of WNB\,0313+683 with its total radio power we can derive an upper limit of 50\degr~ for its orientation with respect to the plane of the sky. 
If the radio axis of the core emission is the same as that of the large scale radio structure, for a projected linear size of this source of 2.0 Mpc we thus find a real linear size $\ga$ 2.6 Mpc.

\begin{table}[tb]
\caption{The measured radio flux densities of the source WNB\,0313+683 and its components.} 
\setlength{\tabcolsep}{1.5mm}
\begin{tabular}{r r @{$\,\pm\,$} l r @{$\,\pm\,$} l r @{$\,\pm\,$} l r @{$\,\pm\,$} l} 
\hline\hline\\
\multicolumn{1}{c}{Freq.} & \multicolumn{2}{c}{Total} & \multicolumn{2}{c}{South Lobe} & \multicolumn{2}{c}{North Lobe} &\multicolumn{2}{c}{Core}\\
\multicolumn{1}{c}{MHz} &\multicolumn{2}{c}{Jy} & \multicolumn{2}{c}{Jy} & \multicolumn{2}{c}{Jy} &\multicolumn{2}{c}{mJy} \\ 
\hline\\
38 & 15.9 & $1.6~^{\rm a}$ \\
151 & 5.4 & $0.5~^{\rm b}$ \\
327 & 3.12 & $0.15~^{\rm c}$ & 2.43 & $0.12~^{\rm c}$ & 0.69 & $0.03~^{\rm c}$ \\
343 & 3.14 & $0.15~^{\rm d}$ & 2.42 & $0.13~^{\rm d}$ & 0.72 & $0.03~^{\rm d}$ \\
1400 & 0.91 & $0.10~^{\rm e}$ \\
1400 & 0.82 & $0.03~^{\rm f}$ & 0.63 & $0.02~^{\rm g}$ & 0.16 & $0.02~^{\rm f}$ & 26 & $2~^{\rm h}$ \\
4850 & 0.34 & $0.04~^{\rm i}$ & 0.24 & $0.03~^{\rm j}$ & 0.05 & $0.01~^{\rm i}$ & 40 & $3~^{\rm k}$ \\
10450 & 0.23 & $0.01~^{\rm l}$ & 0.12 & $0.01~^{\rm l}$ & 0.05 & $0.01~^{\rm l}$ & 55 & $3~^{\rm l}$ \\ 
\ \\
\hline \hline \\
\end{tabular}
{\small Notes: a - 8C survey (Rees 1990, catalogue revised by Hales et al. 1995); b - 6C survey (Hales et al. 1991); c - WENSS ; d - Our 92-cm broadband observations; e - NRAO 1400-MHz survey flux density (Condon \& Broderick 1985) minus NVSS background sources flux; f - NVSS; g - NVSS flux density minus our 1.4-GHz VLA core flux density; h - Our 1.4-GHz VLA observations; i - GB6 survey (Gregory et al. 1996); j - GB6 minus our 4.85-GHz VLA core flux density; k - Our 4.85-GHz VLA observations; l - Our 10.4-GHz Effelsberg observations. The errors in the 38-MHz and 151-MHz points are estimated to be 10\%, the errors in all other observations are the quadratic sum of an estimated 5\% calibration error and a contribution due to noise on the map.}
\label{tab:fluxes}
\end{table}

\subsection{Spectral index maps}

\begin{figure*}
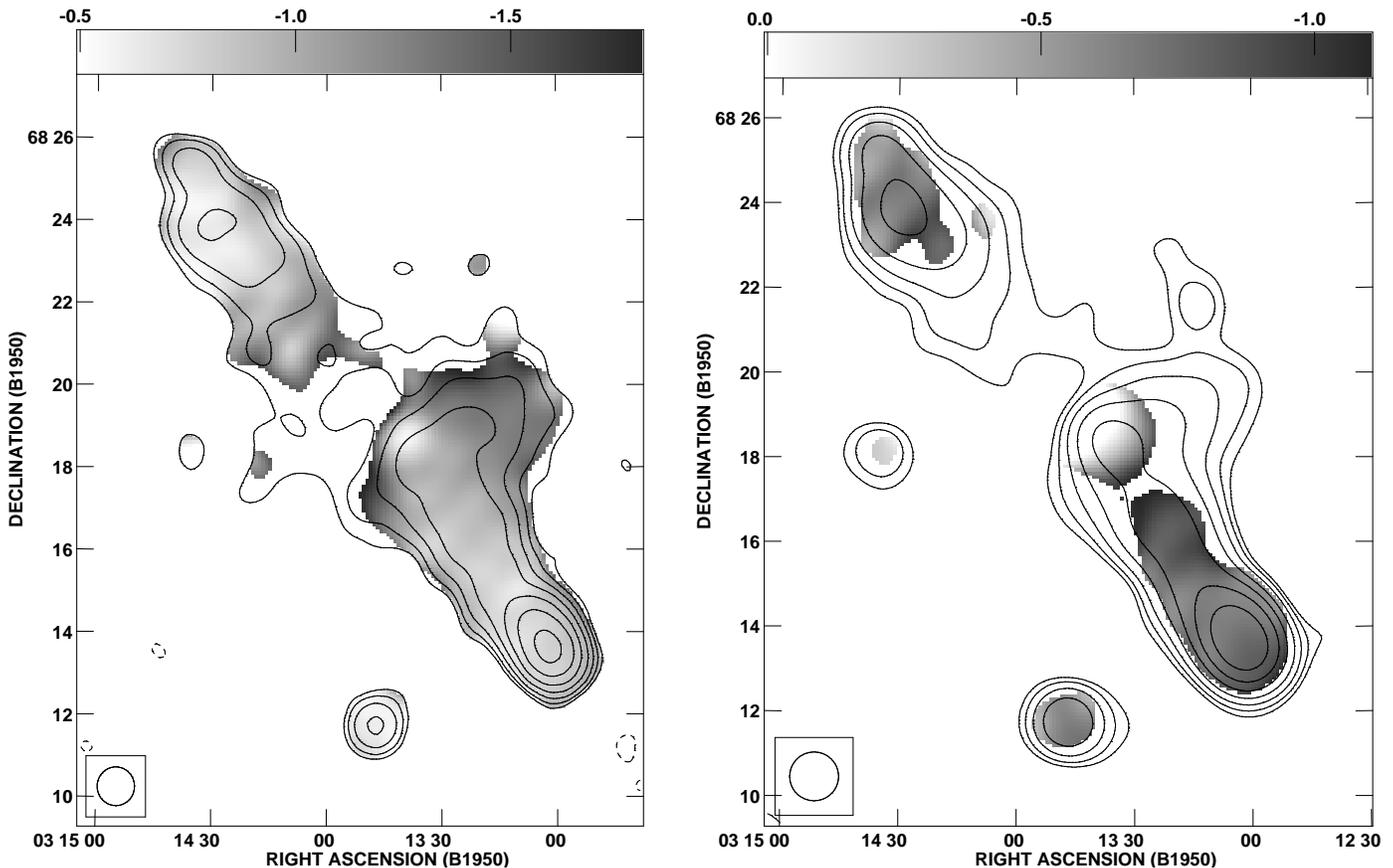

\begin{tabular}{ll}
\psfig{figure=7509.9a,width=0.485\textwidth,clip=} & \psfig{figure=7509.9b,width=0.515\textwidth,clip=}\\
\end{tabular}
\caption{\label{fig:spec_indexmaps}
Greyscale plots of the spectral index distribution in WNB\,0313+683. {\bf a} Spectral index between 327~MHz and 1400~MHz, with contours from the WENSS map. The greyscale ranges from -0.5 (white) to -1.8 (black). {\bf b} Spectral index between between 1400~MHz and 10.45~GHz, with contours from the convolved NVSS map. The greyscale ranges from 0 (white) to -1.1 (black).}
\end{figure*}

We have made spectral index maps of WNB\,0313+683, using the WENSS survey, the NVSS survey, and the Effelsberg observations.     
We decided to use the WENSS survey map rather than the 92-cm broadband WSRT map because it has a much more uniform UV coverage, which is reflected in a well-defined beamshape. This is at the expense of having somewhat lower sensitivity as compared to the 92-cm broadband observations.\\ 
The NVSS survey maps, which are highest in resolution, were convolved to the resolution of either the WENSS survey or the Effelsberg observations. To avoid artefacts, all pixels below $3\sigma_I$ in each map were clipped. The spectral index $\alpha$ was determined for each individual pixel, using the convention that $S_{\nu} \propto \nu^{\alpha}$. The maps are shown in Fig. \ref{fig:spec_indexmaps}.\\
The inverted spectrum core is clearly visible. In both the radio lobes there is a gradual steepening when going inwards from the hotspots towards the core. This is best visible in the low-frequency spectral index map.
Such behaviour is often observed in radio galaxies and can be used to estimate the age of the radio source and the advance velocity of the lobes (e.g. Leahy et al. 1989; Alexander \& Leahy 1987). This will be elaborated further in Sect. \ref{sec:spectral_ages}. The central `bulge' is the region with the steepest spectrum, and therefore most probably contains the electrons with the highest age. The flat-spectrum protrusion at the northern edge of the bulge (near Right Ascension $03^{\rm h}13^{\rm m}13\fs52$, Declination $68\degr21\arcmin38\farcs3$) is due to an unrelated radio source visible in the NVSS maps.\\

\subsection{Polarization properties}
\label{sec:rot_measure_obs}

The mere detection of polarized emission in the 92-cm broadband WSRT observations already gives an upper limit to the Rotation Measure towards WNB\,0313+683. In this observing mode, each channel has a bandwidth of 5~MHz, and the highest frequency channel we used is centered at 355~MHz. Therefore, if $|$RM$| > 60$~rad~m$^{-2}$, the signal within each channel would have been severely depolarized. Both the 92-cm broadband observations and the NVSS detect polarized emission over a large fraction of the southern radio lobe and the central bulge. This gives a total of 6 frequency channels which we can use to find and map the RM distribution over the radio source.\\ 

\subsubsection{Rotation Measures}

\begin{figure*}
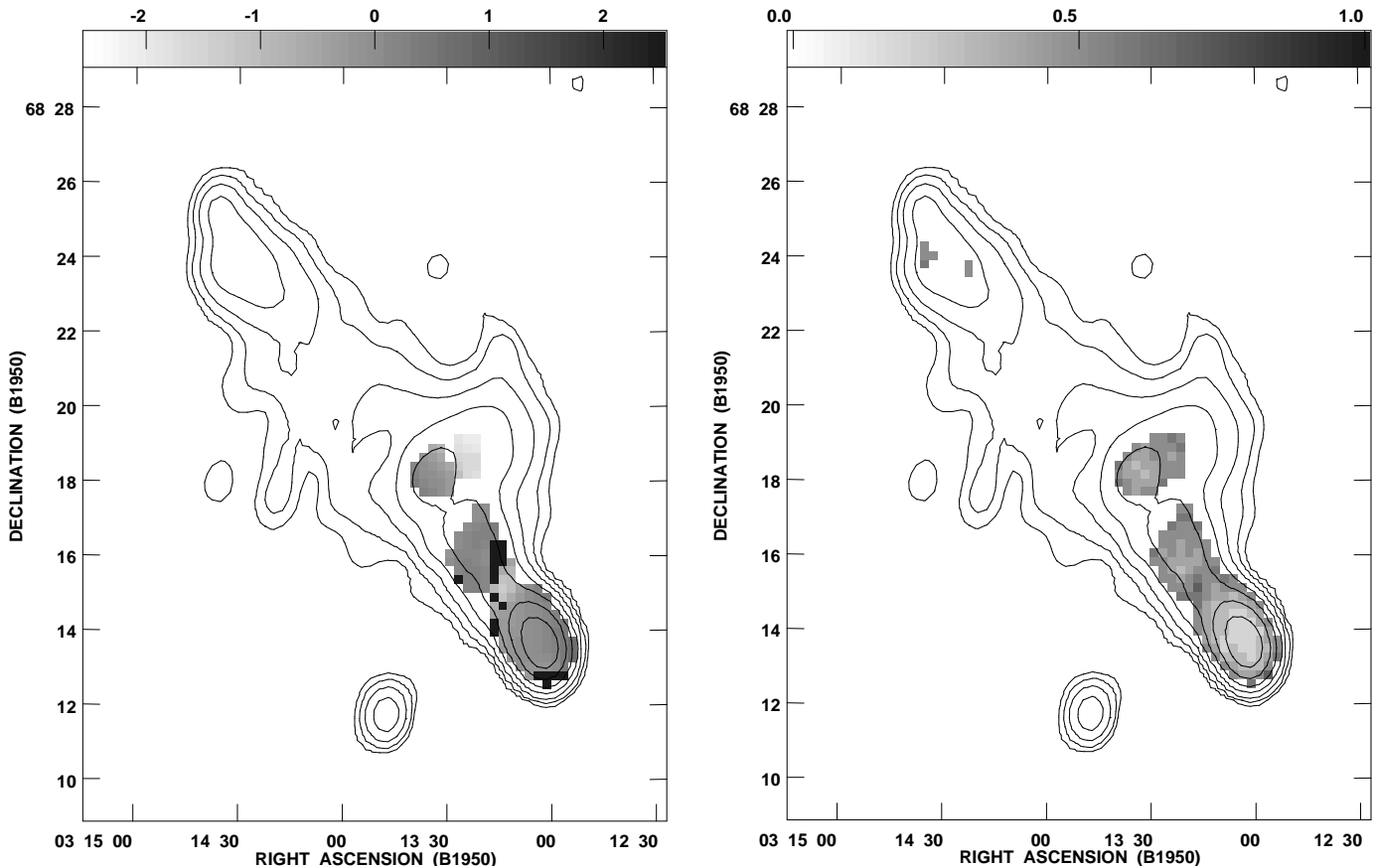

\begin{tabular}{l r}
\psfig{figure=7509.10a,width=0.5\textwidth,clip=} & \psfig{figure=7509.10b,width=0.5\textwidth,clip=} \\
\end{tabular}
\caption{\label{fig:rmmap}{\bf a} Greyscale plot of the Residual Rotation Measure. The Residual Rotation Measure is obtained by adding 10.64~rad~m$^{-2}$ to each pixel. The greyscale ranges from $-2.5$ to $+2.5$~rad~m$^{-2}$. {\bf b} Greyscale plot of the error in the Rotation Measure. The greyscale ranges from 0 to 1~rad~m$^{-2}$. Both maps were obtained using the {\sc newstar} method with the 92-cm WSRT broadband data and the NVSS data combined.}
\end{figure*}

We have used the {\sc newstar} method (see Appendix A) for an analysis of the Rotation Measure distribution over the source. We used the 92-cm broadband and the NVSS data, covering the wavelength range between 20 and 94~cm with 6 channels.  
We have first convolved the NVSS $Q$ and $U$ maps to the resolution of the 92-cm WSRT observations.  
In order to uniformly weight each channel, we have rescaled the NVSS survey $Q$ and $U$ maps with a factor that equalizes the rms noise in the resulting polarized intensity map to that of the low-frequency WSRT polarized intensity maps.\\
On the basis of the simulations shown in the Appendix we have decided to remove all pixels in the averaged polarized intensity maps with values below $10\sigma_{PI}$. Note that even with this rejection level we still expect to find points with a $\sim\!4$~rad~m$^{-2}$ offset with respect to the average value.\\
A map has been made of the Rotation Measure distribution over the source by determining, on a pixel-to-pixel base, the RM at which the averaged polarized intensity is at its maximum.
The RM distribution over the southern radio lobe and the central bulge is very smooth. This is indicative of a galactic origin. 
To find the galactic contribution, we have histogrammed the RMs of the individual pixels, and fitted this histogram with a Gaussian. The mean of the fit was $-10.64$~rad~${\rm m}^{-2}$ with a standard deviation of 0.23~rad~${\rm m}^{-2}$. 
By subtracting this value from the RM-map, we obtain a map of the residual RM, RRM. This map is shown in Fig. \ref{fig:rmmap}, together with a map of the errors in the found RMs. We have defined the error as the half-width of the Rotation Measure profile at a polarized intensity which lies $1\sigma_{PI}$ below the peak of the profile. 
Note that there are indeed a few pixels that have values deviating by $\sim\!4$~rad~m$^{-2}$ from their neighbours, as was predicted by our simulations. We have not corrected for this effect, because the precise value of this deviation is not known for each individual pixel, and because only a few pixels are involved anyway.\\
If the galactic Faraday rotation is uniform over the source, the residual map, as long as its values are significant, indicates structure which is local to the radio source. 
We find that only a small area towards the central bulge has a Rotation Measure which deviates significantly from the mean value, with a RRM of $-1.9 \pm 0.5$~rad~m$^{-2}$. 

\subsubsection{Depolarization}
\label{sec:depol_obs}
\begin{figure*}
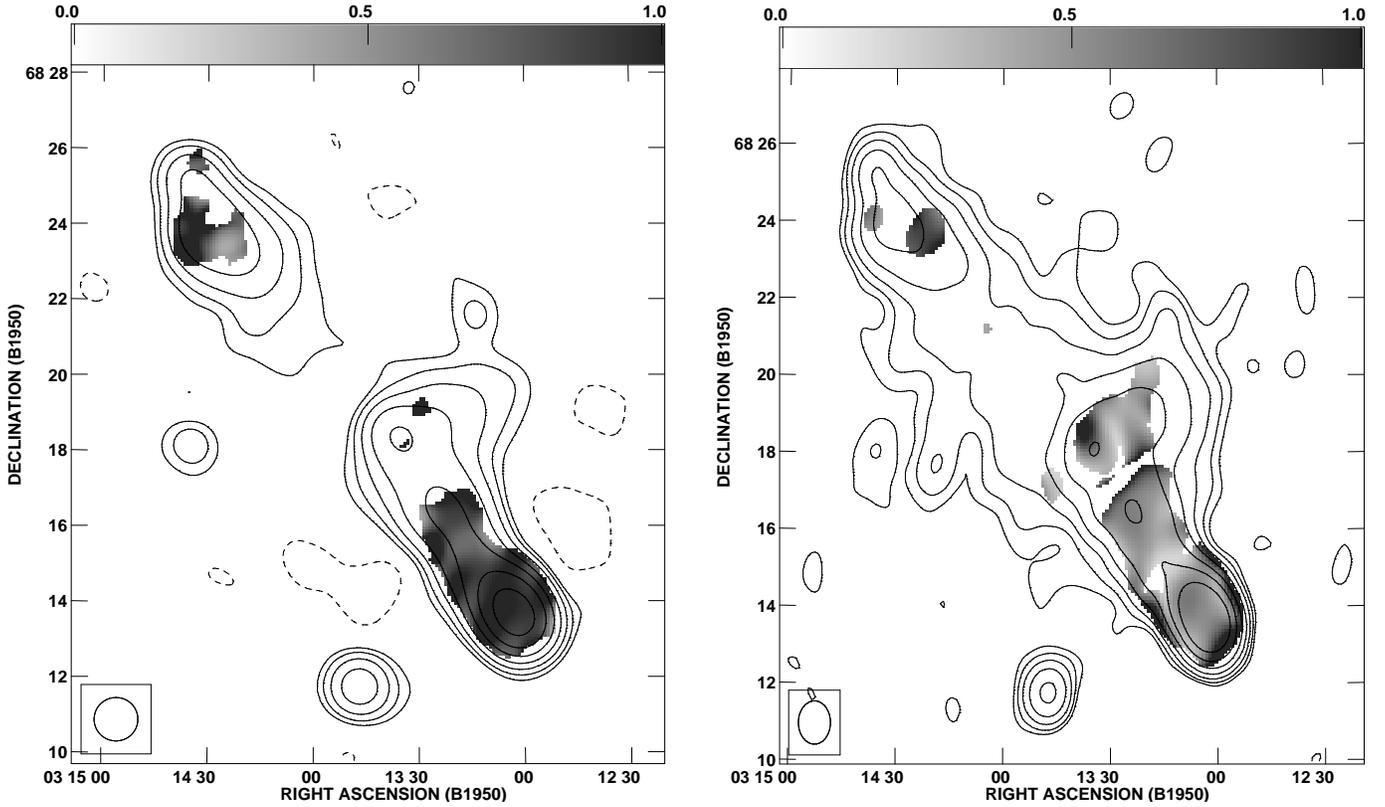

\begin{tabular}{ll}
\psfig{figure=7509.11a,width=0.5\textwidth,clip=} & \psfig{figure=7509.11b,width=0.4932\textwidth,clip=}\\
\end{tabular}
\caption{\label{fig:dp-plots}Plots of the depolarization towards WNB\,0313+683. {\bf a} Depolarization between 10.45~GHz and 1400~MHz, with the depolarization parameter $DP$ ranging from 0 (white) to 1 (black). Contours from the NVSS maps convolved with the Effelsberg beam (69\arcsec~FWHM). {\bf b} Depolarization between 1400 and 343~MHz with $DP$ ranging from 0 (white) to 1 (black). Contours are total intensity from the 92-cm broadband observations.}
\end{figure*}

We have measured the depolarization distribution between the WSRT 92-cm broadband observations and the 1.4-GHz NVSS radio maps.
For the five good channels of the 92-cm broadband observations, maps were made of the polarized intensity $P$ using $P = (Q^2 + U^2 - (1.2\sigma_{QU})^2)^{1/2}$, where $\sigma_{QU}$ is the average of the noise in the $Q$ and $U$ maps. The term containing $\sigma_{QU}$ is a correction for the Ricean bias introduced by the quadratic summation of maps which each have a Gaussian distribution of noise (Wardle \& Kronberg 1974). These five maps were then averaged into one single map.\\
The NVSS $I, Q$, and $U$ maps were smoothed to the resolution of the WSRT 92-cm maps, and maps of $P$ were made.  
All $I$- and $P$-maps were set to zero where the flux in the 92-cm total intensity map did not exceed $3\sigma_I$. Further, pixels in the $P$-maps were blanked when the polarized intensity did not exceed $4\sigma_{QU}$.
At each frequency, maps of the scalar fractional polarization $m' = P/I$ were made (cf. Garrington et al. 1991). From these, the depolarization parameter $DP{_{343}^{1400}} = m'_{343}/m'_{1400}$ was calculated and mapped.\\
To map the depolarization between the 1.4-GHz NVSS and the 10.4-GHz Effelsberg maps, the NVSS maps were convolved with a beam of $69''$~FWHM. The same procedure as above was followed, but using a cut-off of $3\sigma_I$ in the NVSS maps. 
Figure \ref{fig:dp-plots} shows the greyscale plots of the depolarization between 343 and 1400~MHz, and between 1400 and 10450~MHz. At the higher frequencies we see only marginal depolarization ($DP{^{10450}_{1400}} \approx 1$), but at the lower frequencies the depolarization is much stronger.\\
Towards the radio core we measure $DP{_{343}^{1400}}>1$. This must be due to the increased contribution of the unpolarized radio core to the total intensity at 1400~MHz. It causes a decrease of the fractional polarization and thus increases the value of $DP{_{343}^{1400}}$.

\subsection{Physical properties of the radio source}

\label{sec:phys_prop}
We used the measured flux densities at 151 and 327~MHz to interpolate the flux density at 178~MHz. Using this flux density and the redshift of WNB\,0313+683, we derive a radio power at 178~MHz of $P_{178} = 1.71 \times 10^{26}$~W~Hz$^{-1}$. This power is comparable to that which distinguishes FRI-type from FRII-type radio sources (Fanaroff \& Riley 1974), like in other Mpc-sized radio sources (e.g. Saripalli et al. 1986).\\
We have calculated the equipartition values of the energy density and the magnetic field strengths of the entire source, the two lobes, and along two slices roughly halfway between the radio core and the hotspots in each lobe. We use the standard method outlined in Miley (1980), but using a conical, rather than cylindrical, geometry for each side of the source because of the existence of the central bulge. This gives a correction factor of $\frac{1}{3}$ in the volume calculation. We use a base diameter of 840~kpc, which is the diameter of the central bulge measured between the $6\sigma_I$ contours on the 92-cm WSRT map, and a length of 2000~kpc, i.e. the projected length.
We assume an equal amount of energy in relativistic electrons and protons ($k=1$), and a filling factor of 1. 
We use the integrated flux densities at 343~MHz of the 92-cm broadband observations. The results are shown in Tab. \ref{tab:energies}.\\ 
From the energy densities the equipartition pressures have been calculated assuming that the pressure is dominated by relativistic particles. In that case the pressure is just $\frac{1}{3}\,u_{eq}$. The energy densities and pressures of the lobes are best represented by the values resulting from the slices, since they do not incorporate the bright and compact hotspots. The equipartition energy densities and pressures we find are well within the range found in other Mpc-sized radio galaxies (Subrahmanyan \& Saripalli 1993, Mack et al. 1998).
We also find low equipartition magnetic field strengths, $\sim 5 \times 10^{-7}$~G. Since, at a redshift of 0.0901, the equivalent magnetic field strength of IC scattering of Microwave Background photons is $\sim 3.9 \mu$G, this must be the dominant energy loss mechanism for the relativistic electrons in the lobes (see Sect. \ref{sec:spectral_ages} for a further discussion of this topic).\\ 

\begin{table}
\caption{\label{tab:energies}Equipartition parameters of WNB\,0313+683. All values are calculated assuming equal energies in electrons and protons and a filling factor of 1. The energy density is denoted by $u_{eq}$, the pressure by $p_{eq}$ and is just $\frac{1}{3}u_{eq}$. $B_{eq}$ is the equipartition magnetic field strength. These three parameters are calculated for the source as a whole, for the two radio lobes, and for two slices perpendicular to the radio axis located halfway between the core and the hotspot.}
\begin{tabular}{lccc}
\hline \hline \\
Component & $u_{eq}$ & $p_{eq}$ & $B_{eq}$ \\
 & $10^{-14}$~erg~cm$^{-3}$ & $10^{-14}$~dyn~cm$^{-2}$ & $\mu$G \\
\hline \\
Total & \phantom{1}7.7 & 2.6 & 0.9 \\
North lobe & \phantom{1}4.5 & 1.5 & 0.7 \\
North slice & \phantom{1}1.2 & 0.4 & 0.4 \\
South lobe & 11.0 & 3.7 & 1.1 \\
South slice & \phantom{1}1.9 & 0.6 & 0.5 \\
\ \\
\hline \hline \\
\end{tabular}
\end{table}

\subsection{Physical properties of the AGN from the optical spectrum}

\subsubsection{Extinction determination}

Table \ref{tab:linefluxes} lists the measured line strengths of the optical spectrum of the host galaxy of WNB\,0313+683.
From the ratio of the H$\alpha$ and H$\beta$ line strengths we have determined the color excess $E(B-V)$ to be $0.90 \pm 0.10$~mag if it is
local to the source, or $0.98 \pm 0.10$~mag if it is caused in our own
galaxy. We used the interstellar extinction curve from Cardelli et al. (1989), and an unreddened H$\alpha$/H$\beta$ ratio of 3.1, as suggested by Halpern \& Steiner (1983) and Gaskell \& Ferland (1984).
Since we do not detect a broad component of the H$\beta$ line, we only use the narrow line components.\\
The galactic latitude of WNB\,0313+683 is $+9\fdg8$. From the Leiden-Dwingeloo HI-survey (Hartmann 1994) we have extracted the galactic HI column density towards the radio source. We find a column density $N(HI)$ of $2.86 \times 10^{21}$~atoms~cm$^{-2}$, integrated from $-450$ km~s$^{-1}$ to $+400$ km~s$^{-1}$. Using a conversion factor of $N(HI)/E(B-V) = 5.6 \times 10^{21}$~atoms~cm$^{-2}$~mag$^{-1}$ (Burstein \& Heiles 1978), we estimate the galactic color excess $E(B-V) = 0.51$~mag. We note however that the HI column density map of this region is highly structured, so that higher density patches may exist within a single beam of the Leiden-Dwingeloo survey ($\sim 30\arcmin$). 
Most of the observed extinction is thus probably galactic in origin, and we have corrected the line fluxes using an $E(B-V)$ of $0.98$~mag. These corrected line strengths are also printed in Tab. \ref{tab:linefluxes}.

\subsubsection{Physical properties of the emission-line region}

\begin{figure}
\psfig{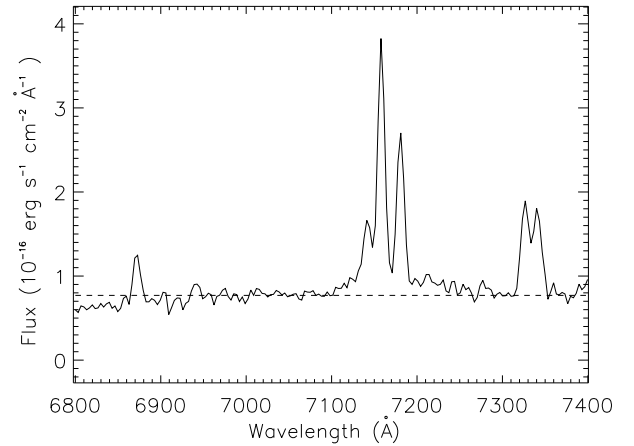}
\caption{Enlargement of the optical spectrum between 6800\AA~and 7400\AA, containing the $[$OI$]$, H$\alpha$+$[$NII$]$ and $[$SII$]$ emission lines. The dashed line roughly indicates the level of the continuum around the H$\alpha$+$[$NII$]$ lines}
\label{fig:ha-detail}
\end{figure}

Figure \ref{fig:ha-detail} shows the spectrum in the range surrounding the H$\alpha$ line. Clearly, the line profile of H$\alpha$ has a broad base. The faint broad wings of H$\alpha$ can be seen to extend from roughly 7100\AA~to 7250\AA, equivalent to $\sim 6000$~km~s$^{-1}$. H$\alpha$ is the only permitted line with a broad component in our spectrum. WNB\,0313+683 is not exceptional in this case. Some other GRGs have been found to possess broad lines, such as WNB\,1626+5152 (R\"{o}ttgering et al. 1996) and 0319+411 (de Bruyn 1989).\\ 
From the corrected line strengths we can derive the temperature and density of the narrow-line emitting clouds. From the ratio of the $[$OIII$]$ lines, $([$OIII$]4959 + 5007)$ / $[$OIII$]4363$, we find a temperature in the line emitting regions of $1.6^{+0.8}_{-0.3} \times 10^4$~K. The large error is mainly due to the large uncertainty in the line flux of $[$OIII$]4363$. 
The ratio of the amount of flux in the $[$SII$]6717,6734$ emission lines gives an electron density ${\rm n}_{\rm e}$ of $7^{+4}_{-2}\times10^2$~cm$^{-3}$ in the line-emitting regions.

\begin{table*}[tb]
\caption{\label{tab:linefluxes}Measured lines and fluxes in the spectrum of WNB\,0313+683. The fluxes are measured using Gaussian fits to the observed line profiles. The numbers between the brackets denote the errors in the measured quantities. For the H$\alpha$/$[$NII$]$ complex, the lines were deblended using three Gaussians; for the $[$SII$]$ lines, two Gaussians were used. The powers have been calculated using a mean redshift of $0.0901$, and assuming isotropic emission.}   
\begin{tabular}{lccc r@{\ }l r@{\ }l r@{\ }l r@{\ }l}
\hline \hline\\
 & & & &\multicolumn{2}{c}{Measured} & & & \multicolumn{4}{c}{Extinction corrected$^a$} \\
Line & $\lambda_{\rm obs}$ & $z$ & FWHM~$^b$ & \multicolumn{2}{c}{Flux~$^b$} & \multicolumn{2}{c}{EW} & \multicolumn{2}{c}{Flux~$^c$} & \multicolumn{2}{c}{Power~$^c$} \\
 & \AA & & \multicolumn{1}{c}{\AA} & \multicolumn{2}{r}{\small $10^{-15}$~erg~s$^{-1}$~cm$^{-2}$} & \multicolumn{2}{c}{\AA} & \multicolumn{2}{l}{\small $10^{-15}$~erg~s$^{-1}$~cm$^{-2}$} & \multicolumn{2}{c}{\small $10^{41}$~erg~s$^{-1}$}\\
\hline \\
$[$OII$]$3727      & 4062.0 & 0.0899 & \phantom{1}6.7 & \hspace{0.5cm} 0.56 & (0.08) & 169\phantom{.} & (50)   & \hspace{0.5cm} 32.2 & (4.6) & \hspace{0.1cm} 44.3 & (6.3)\\
H$\gamma$          & 4730.3 & 0.0902 &  \phantom{1}7.3 & 0.15 & (0.04) &  7.2\phantom{1} & (2.0) &  4.5 & (1.2) &  6.0 & (1.5)\\
$[$OIII$]$4363     & 4755.7 & 0.0900 &  \phantom{1}6.2 & 0.08 & (0.05) &   4.4\phantom{1} & (1.0) &  2.1 & (1.2) &  2.7 & (1.4)\\
H$\beta$           & 5299.4 & 0.0902 &  \phantom{1}9.2 & 0.41 & (0.04) &   2.4\phantom{1} & (1.5) &  7.7 & (0.8) & 10.5 & (3.0)\\
$[$OIII$]$4959     & 5406.2 & 0.0902 &  \phantom{1}9.3 & 1.36 & (0.04) &  41.3 & (4.2) & 23.8 & (0.7) & 32.6 & (1.0)\\
$[$OIII$]$5007~$^d$& 5458.1 & 0.0901 &  \phantom{1}8.6 & 4.10 & (0.04) & 126\phantom{.} & (13)   & 69.6 & (0.7) & 95.1 & (1.0)\\
$[$OI$]$6300       & 6869.6 & 0.0904 &  \phantom{1}9.3 & 0.59 & (0.04) &   9.0\phantom{1} & (1.0) &  5.2 & (0.4) &  7.1 & (0.4)\\
$[$OI$]$6364       & 6939.1 & 0.0904 &  \phantom{1}9.4 & 0.22 & (0.05) &   3.0\phantom{1}& (0.4) &  2.0 & (0.4) &  2.6 & (0.4)\\
$[$NII$]$6548~$^e$ & 7139.1 & 0.0903 &  \phantom{1}9.1 & 0.82 & (0.2)  &   8.7\phantom{1}& (3.0) &  6.3 & (0.8) &  8.7 & (0.4)\\
H$\alpha$~$^e$     & 7156.0 & 0.0904 &  \phantom{1}9.1 & 3.10 & (0.2)  &  32.8 & (3.0) & 23.9 & (0.8) & 32.7 & (0.3)\\
$[$NII$]$6583~$^e$ & 7178.3 & 0.0904 &  \phantom{1}8.7 & 1.87 & (0.2)  &  19.8 & (3.0) & 14.3  & (0.8) & 19.6 & (0.2)\\   
$[$NII$]$ + H$\alpha$~$^f$ & &       &     & 7.9\phantom{1}  & (0.4)  & 103\phantom{.} & (10)   & 60.8 & (3.1) & 84.6 & (4.3)\\
$[$SII$]$6717      & 7323.9 & 0.0904 &  \phantom{1}9.7 & 1.23 & (0.04) &  15.9 & (2.0) &  8.8 & (0.3) & 12.3 & (0.4)\\
$[$SII$]$6734      & 7338.6 & 0.0900 & 10.8 & 1.25 & (0.04) &  16.2 & (2.0) &  8.9 & (0.3) & 12.5 & (0.4)\\
$[$SII$]$ total    &        &        &     & 2.51 & (0.17) &  33.4 & (3.0) & 18.2 & (1.2) & 25.3 & (1.7)\\
\ \\
\hline \hline \\
\end{tabular}
\ \\
{Notes:}\\
{$^a$~Using an $E(B-V)$ of $0.98$ mag. (galactic extinction derived using an H$\alpha/$H$\beta$ line-ratio of $3.1$).}\\
{$^b$~Measured using Gaussian fits to the line profiles.}\\
{$^c$~The errors do not incorporate the error of 0.10 in the color index; when included, this leads to an additional error of 30\% in the corrected flux and power.}\\
{$^d$~Overlaps with the HgI atmospheric line at 5461\AA}\\
{$^e$~Values are calculated using a 3-component Gaussian fit to the (blended) lines. The given H$\alpha$ flux is for the narrow component only.}\\ 
{$^f$~These values include the broad component of H$\alpha$. The total flux in H$\alpha$ is $5.17 \times 10^{-15}$~erg~s$^{-1}$~cm$^{-2}$.}
\end{table*}

\subsubsection{Optical emission line luminosity and jet power}
\label{sec:oell}
Rawlings \& Saunders (1991) report a positive correlation between the optical emission line luminosity (OELL) of radio galaxies and quasars, and their jet power. Although the spread in this correlation is large, it holds over several orders of magnitude. 
In their sample they included several known GRGs, and they all fit well in this relation.\\
In Sect. \ref{sec:density_ram} we derive that the total jet power is $4.0\times10^{44}$~erg~s$^{-1}$. The method we use to calculate this is similar to that of Rawlings \& Saunders, so the results are directly comparable.\\
To estimate the total luminosity in optical emission lines we also use the same method as Rawlings \& Saunders: $L_{tot} = 3\times(3\times L_{[OII]3727} + 1.5 \times L_{[OIII]4959 + 5007})$. From Tab. \ref{tab:linefluxes} we find $L_{tot}  = (9.7 \pm 0.7) \times 10^{43}$~erg~s$^{-1}$.
When we compare WNB\,0313+683 with the sources of Rawlings \& Saunders, we find that it lies well away from their correlation: Its OELL is a factor 10--20 too high for its jet power (or, equivalently, its jet power is a factor 10--20 too low for its emission-line power). We note, however, that when we use a galactic color excess $E(B-V)$ of 0.51 mag., as predicted by the HI column density from Hartmann (1994), WNB\,0313+683 agrees much better with the correlation of Rawling \& Saunders, although the OELL still is on the high side.  

\section{Analysis of the spectral index distribution}

\subsection{Spectral index profiles}

We have integrated the total intensity distribution of WNB\,0313+683 in boxes perpendicular to the radio axis at 327, 1400, and 10450~MHz. The NVSS 1400-MHz data were first convolved to the WENSS or Effelsberg resolution. The width of the slices is half a beamwidth (i.e. 30\arcsec~at 327~MHz, 35\arcsec~at 10.45~GHz), their length is $\sim 200\arcsec\,(\sim 500$~kpc).\\
Since we are primarily interested in the behaviour of the spectral index as a function of distance from the hotspots, we have plotted this for each lobe separately. 
The positions of the hotspots have been taken from the high-resolution VLA observations. The results are shown in Fig. \ref{fig:ages}.\\
Between 327 and 1400~MHz, both lobes show a significant steepening of their spectrum with increasing distance from the hotspot. However, both lobes also show a small but significant {\em flattening} in the first 100--200~kpc behind the hotspots. In the southern lobe, the peak coincides roughly with the secondary hotspot that is well visible in the VLA maps, but in the northern hotspot there is no such counterpart. 
The spectrum near the secondary hotspot in the southern lobe is flatter between 1400 and 10450~MHz than between 327 and 1400~MHz. 
Because of the faint high-frequency emission of the bridge of the northern lobe a spectral index could be accurately measured in this region. The strong increase at $\sim 800$~kpc in the southern lobe and at $\sim 1100$~kpc in the northern lobe is due to the inverted spectrum radio core.

\subsection{Expansion velocity, backflows, and the age of WNB\,0313+683}
\label{sec:spectral_ages}

\begin{figure*}
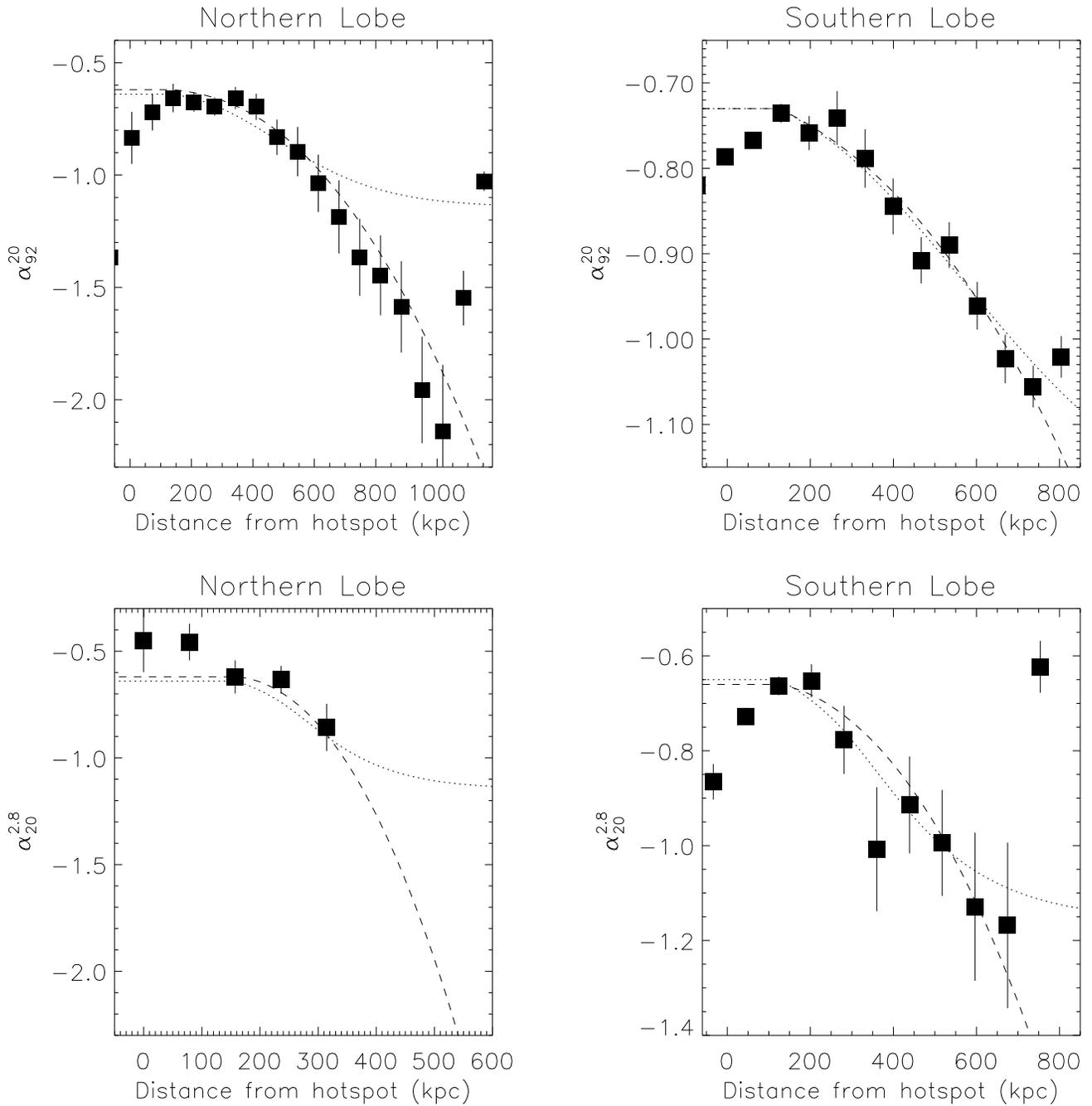

\begin{tabular}{ll}
\psfig{figure=7509.13a,width=0.5\textwidth} & \psfig{figure=7509.13b,width=0.5\textwidth} \\
\psfig{figure=7509.13c,width=0.5\textwidth} & \psfig{figure=7509.13d,width=0.5\textwidth}\\
\end{tabular} 
\caption{Plots showing the spectral index profiles along the radio axis of the two lobes of WNB\,0313+683. The two upper panels show the spectral index profiles between 327 and 1400~MHz, $\alpha_{92}^{20}$, the lower panels between 1400~MHz and 10.45~GHz, $\alpha_{20}^{2.8}$. Because of the lower resolution of the Effelsberg data, the sizes of the bins are somewhat larger in the high-frequency plots. Also plotted are the best fitting profiles from the model fits (see text for details). The dashed lines indicate the fits using a JP, the dotted lines using a CI model. \label{fig:ages}}
\end{figure*}

The spectral index profiles have been used to estimate the age of the radio source WNB\,0313+683 and the expansion velocities of the two radio lobes. Ageing of a population of relativistic electrons results in a steepening of the emitted radio spectrum above a certain frequency (e.g. Kardashev 1962, Pacholzcyk 1970).
The three standard models that describe the change in the spectrum are the Kardashev-Pacholzcyk (KP), the Jaffe-Perola (JP) and the Continuous Injection (CI) model; see Carilli et al. (1991) for an excellent overview of these models. The CI model incorporates a continuous injection of `fresh' electrons, the JP and KP models do not and are therefore `pure' ageing models. 
This, in principle, makes the CI model more suited to describe integrated spectra of radio sources, whereas the KP and JP models are more suited for analysing spectra of source components which do not contain possible sources of acceleration.
In terms of the spectral shape, the JP and the CI model give the most, respectively the least amount of steepening. Therefore we decided to use only these two models in the following analysis.\\
The important parameters that describe the evolution of an ensemble of radiating electrons are the time past since their last reacceleration, the magnetic field strength, the intensity of the microwave background radiation and the low-frequency spectral index of the radiation, often called the injection spectral index $\alpha_{inj}$. 
Other factors also modify the radio spectrum, such as expansion of the radio emitting plasma. However, our analysis concentrates on the bridges of the radio source, where we believe expansion is not important any more.\\
The equivalent field strength of the microwave background radiation (MWBR) is given by $3.24\,(1+z)^2~\mu$G, adopting a present-day temperature of 2.726~K for the MWB radiation. At the redshift of WNB\,0313+683 ($z = 0.0901$) this yields $B_m = 3.9 \mu$G, which is almost ten times larger than the internal equipartition magnetic field strength in the radio lobes (see Tab. \ref{tab:energies}). Therefore IC scattering of the MWBR photons must be the dominant energy loss factor for the electrons in the lobes of WNB\,0313+683.
The maximum amount of time past since the last acceleration of the electrons, which we define as the age of the source, is then obtained by assuming that the internal magnetic field strength of the source $B_s = B_m/\sqrt{3}$ (e.g. van der Laan \& Perola 1969). For WNB\,0313+683, an upper limit on the source age is thus obtained when $B_s = 2.1 \mu$G. We have used this value in our spectral age calculations.\\
To find the advance velocities of the hotspots and the ages of the lobes we used the following method. Under the assumptions given above, we have first calculated a library of radio spectra as a function of age of the source and injection spectral index. From these we have derived tables of the spectral indices between 325 and 1400 MHz, and between 1400 MHz and 10.45 GHz.\\
For a fixed advance velocity of the hotspot and $\alpha_{inj}$, each position in the radio lobe has a unique age (neglecting possible backflows) and thus a unique spectral index between two fixed frequencies. 
So, using a range of $\alpha_{inj}$ and advance velocities, we can compare the expected spectral index behaviour with our observations. How well the expected behaviour fits the observations is measured by calculating the (reduced) $\chi^2_{\nu}$. This results in an array of $\chi^2_{\nu}$ values. In this array we search for the minimum $\chi^2_{\nu}$, which gives the parameters that fit the observations best. 
The errors in the fitted parameters are found by searching the 1$\sigma$ error-ellipse in the $\chi^2_{\nu}$-array, which for a two-parameter problem is outlined by array positions with a value of the minimum $\chi^2_{\nu}$ plus 2.3 (e.g. Wall 1996). This procedure has been followed for both JP and CI models.\\ 
Our method differs from the more conventional way of measuring advance velocities of hotspots from two-frequency radio data (e.g. Myers \& Spangler 1985). Usually, one assigns a certain $\alpha_{inj}$, often taken as the hotspot spectral index, and then calculates the age of each measured point in the lobe. When this is plotted against hotspot distance a linear fit is made through the measured points in order to obtain a velocity. The advantage of our method is that both $\alpha_{inj}$ and the hotspot velocity are derived from the data, and that reliable error estimates for both values are obtained. The minimizing of the $\chi^2_{\nu}$ ensures that the best solution under the given assumptions is found.\\
One important point to take into consideration is that our method only gives physically relevant results if the position of the last acceleration of the electrons is known. This is most likely the position near the head of the lobe with the flattest spectrum, which usually coincides with the primary hotspot.  
In the southern lobe of WNB\,0313+683 the secondary hotspot has the flattest spectrum between 327 and 1400~MHz. We have therefore used the position of the secondary hotspot as the zero-age point for the radiating electrons. 
Also in the northern lobe the region with the flattest spectrum does not lie at the head of the lobe. Again, we have decided to use the point with the flattest spectrum as the zero-age point.\\ 
In the southern lobe, we have fitted both the low- and high-frequency spectral index profiles to find out if they yield consistent results. In the northern lobe, we only used the profile of the spectral index between 327 and 1400~MHz. The 10.45-GHz flux densities are too low to be used for any fitting.
However, as a consistency check, we have plotted the profiles for the expected spectral index distribution between 1400~MHz and 10.45~GHz.\\ 
The best fitting values for $\alpha_{inj}$ and the hotspot velocity have been presented in Tab. \ref{tab:specfit}. Note that the errors given in the table are only the formal errors resulting from the fit procedure, and do not take into account possible systematical errors in the data. The spectral index profiles resulting from these parameters have been drawn on top of the measured spectral index profiles (Fig. \ref{fig:ages}).\\
We find that the best fit of the low-frequency spectral index profile in the northern lobe is yielded by a JP model, with a lobe advance velocity of $0.027 \pm 0.002c$. The high-frequency profile constructed with this velocity is not inconsistent with the measured data points.
For the southern lobe, the JP and CI model fit the spectral index profile equally well, both at low and high frequencies. Since the low-frequency spectral index measurements have smaller errors, we will use those results for an upper and lower limit on the hotspot advance velocity in the southern lobe.\\
The good fit of the JP model to the northern lobe indicates that the radiating particles in that lobe must be relatively undisturbed. The southern lobe, with its much higher surface brightness, brighter and double hotspot, and its central bulge, is much more indicative of a backflow with mixing and re-acceleration of the radiating particles. This might be why the CI model still provides a good fit to the data here.\\   
If these results have any physical meaning (which, because of the large number of assumptions, is questionable; see e.g. Eilek 1996), the advance velocities of the hotspots of WNB\,0313+683 are between 0.02 and 0.04$c$.
Velocities of radio lobes have been determined for a sample of powerful 3C sources by Alexander \& Leahy (1987). They find velocities in the range of 0.01$c$ - 0.2$c$, and a weak correlation of lobe velocity with radio power. WNB\,0313+683 agrees quite well with that correlation, since it has a low radio power and a low lobe advance velocity. The same was found for other GRGs (e.g. Parma et al. 1996, Lacy et al. 1993).\\
The age of the northern lobe can be estimated by calculating the crossing time from the core to the head of the lobe ($\sim 1200$~kpc) with a velocity of $0.027 \pm 0.002c$. This results in an age of $1.4 \pm 0.1 \times10^8$~yrs. Similarly, for the southern lobe, we find an age between $0.7\pm 0.1 \times 10^8$~yrs (JP model) and $1.4\pm 0.1 \times 10^8$~yrs (CI model). If backflows in the lobes are important, then this will increase the estimated age of the source.\\ 
The assumptions we have made were meant to provide us with an upper limit on the source age. We therefore use the highest age we find, $1.4 \pm 0.1 \times10^8$~yrs, as an upper limit.
Ages around $10^8$ yrs have been found to be fairly typical of GRGs (e.g. Mack et al. 1998).

\begin{table}[tb]
\caption{\label{tab:specfit}The fitted injection spectral index $\alpha_{inj}$ and the advance velocities of the northern and southern radio lobe. For the southern lobe, both the low and high frequency spectral index profiles were fitted. Velocities are given in units of the speed of light $c$. Also given is
the reduced $\chi^2$ of the fit. The last column gives the range of points which were used for the fit, in kpc from the hotspot.}
\setlength{\tabcolsep}{1.5mm}
\begin{tabular}{l r @{$\,\pm\,$} l r @{$\,\pm\,$} l r c c }
\hline \hline \\
Model & \multicolumn{2}{c}{$\alpha_{inj}$} & \multicolumn{2}{c}{$v/c$} & $\chi_{\nu}^2$ & Fit range \\
\hline \\
\multicolumn{7}{c}{\bf Southern Lobe}\\
\ \\
\multicolumn{7}{c}{Low frequency spectral indices} \\
CI & $-0.73$ & 0.02 & 0.020 & 0.002 & 0.65 & 130 - 700 \\
JP & $-0.73$ & 0.02 & 0.040 & 0.003 & 0.63 &  \\
\ \\
\multicolumn{7}{c}{High frequency spectral indices} \\
CI & $-0.65$ & 0.02 & 0.024 & 0.005 & 0.73 & 100 - 600 \\
JP & $-0.66$ & 0.02 & 0.060 & 0.005 & 0.77 &  \\
\ \\
\hline \\ 
\multicolumn{7}{c}{\bf Northern Lobe}\\
\ \\
\multicolumn{7}{c}{Low frequency spectral indices} \\
CI & $-0.64$ & 0.03 & 0.014 & 0.002 & 3.78 & 100 - 1000 \\
JP & $-0.62$ & 0.03 & 0.027 & 0.002 & 0.38 &  \\
\ \\
\hline \hline \\
\end{tabular}
\end{table}

\subsection{The density around the radio lobes (from ram-pressure arguments)}
\label{sec:density_ram}
In case that the head of a lobe is ram-pressure confined, the density $\rho_a$ of the ambient medium can be found using $\rho_a = \Pi_j / (A_h v_h^2)$, where $\Pi_j$ is the thrust of the jet, $A_h$ is the area of the bowshock and $v_h$ is the advance velocity of the head of the lobe.
The thrust $\Pi_j$ of the jet is given by  $Q_{jet} / v_j$ (e.g. Begelman et al. 1984, their Appendix B.3), with $v_j$ the velocity of the material in the jet and $Q_{jet}$ the amount of energy delivered by the jet per unit time, or the jet power.
We can estimate this by dividing the total energy content of the radio source by the age of the source.\\
The energy density of WNB\,0313+683 has been measured and is given in Tab. \ref{tab:energies} as $7.7 \times 10^{-14}$~erg~cm$^{-3}$. To estimate the volume occupied by the radio lobes, we again assume that WNB\,0313+683 has a double conical morphology (because of the central bulge), with a total length of
2000~kpc and a base diameter of 840~kpc (6$\sigma$ contours in the 92-cm WSRT radiomap). We thus find a total volume of $3.7 \times 10^8$~kpc$^3$. Therefore the total energy content of the radio source, assuming it is in equipartition, is $8.4\times10^{59}$~erg.\\ 
An upper limit for the age of the source has been found to be $1.4 \times 10^8$~yrs. If the jet power $Q_{jet}$ has been constant during this time, the energy delivered by each of the two jets is $Q_{jet} \ga \eta^{-1} \times 10^{44}$~erg~s$^{-1}$. Here $\eta$ is the efficiency by which the kinematic energy carried by the jet is converted into radiation. 
A conservative choice of $\eta$ is given by Rawlings \& Saunders (1991), who take $\eta = \frac{1}{2}$ stating that half of the energy is lost to work done by expansion and half of the energy is transferred into radiation. 
Norman et al. (1982) have found that $\eta$ depends on the density contrast between the material in the jet and the external medium. Since the density of the external medium of GRGs is low, the efficiency is probably also low unless the density of the material in the jet is low as well. As yet, there are no observational constraints on the efficiency, so we will use the conservative choice $\eta = \frac{1}{2}$ to find a lower limit on the jet-power. 
This results for WNB\,0313+683 in $Q_{jet} \ga 2 \times 10^{44}$~erg~s$^{-1}$. The velocity of the material flowing down the jet we set at $c$, the speed of light, to obtain a lower limit on the thrust.\\
The area of the bowshock is difficult to find since this is not directly observable. If we assume that it is not larger than the area of the observed hotspots, we can use our VLA observations to find an upper limit. To obtain maximal spatial resolution, we made new maps of the VLA data using uniform, rather than natural weighting of the UV-data. We then fitted both hotspots with a single Gaussian, using the {\sc jmfit} program in the {\sc AIPS} software package.
The deconvolved diameter of the southern hotspot is 8\farcs5, which translates into a maximum impact area of 280~kpc$^2$. For the advance velocity of the hotspot we use $0.03c$, which is the mean of the two values from Tab. \ref{tab:specfit}. We thus find a lower limit on the density of the external medium $\rho_a \ga 3.7 \times 10^{-30}$~g~cm$^{-3}$, or on the particle density $n_a \ga 1.6 \times 10^{-6}$~cm$^{-3}$ (using a mean atomic mass per particle of 1.4 amu).\\
The hotspot in the northern lobe has a deconvolved width of 14\arcsec, equivalent to an impact area of 780~kpc$^2$.
Using an advance velocity of $0.027c$, we find an ambient density $\rho_a \ga 1.5 \times 10^{-30}$~g~cm$^{-3}$ and for the particle density $n_a \ga 5.8 \times 10^{-7}$~cm$^{-3}$.\\
The densities we find are in good agreement with those found around other Mpc-sized radio galaxies (e.g. Saripalli et al. 1986, Parma et al. 1996, Mack et al. 1998). 
The density contrast between the two sides coincides with the asymmetry in the hotspot flux: the side with the higher luminosity is also the side with the higher external density. This agrees with the prediction that the efficiency of the acceleration process increases with increasing density of the ambient medium (e.g. Norman et al. 1982).\\
We have so far assumed that the jet is a straight particle stream. In case that it precesses, the area of the bowshock in the above analysis should be replaced by the much larger cross-section of the entire radio lobe (Scheuer 1982). This results in much lower external densities ($\sim\!8 \times 10^{-9}$~cm$^{-3}$ for the southern lobe). An indication for the presence of such a precession is given by the observation of two hotspots in the southern lobe. Numerical simulations of precessing jets reproduce such structures very well (e.g. Cox et al. 1991). A valuable, but difficult, observation would be to map the jet along the entire length of the lobe.

\section{The origin of the observed depolarization}
\label{sec:depol_analysis}

\subsection{The Faraday dispersion}

\begin{figure*}
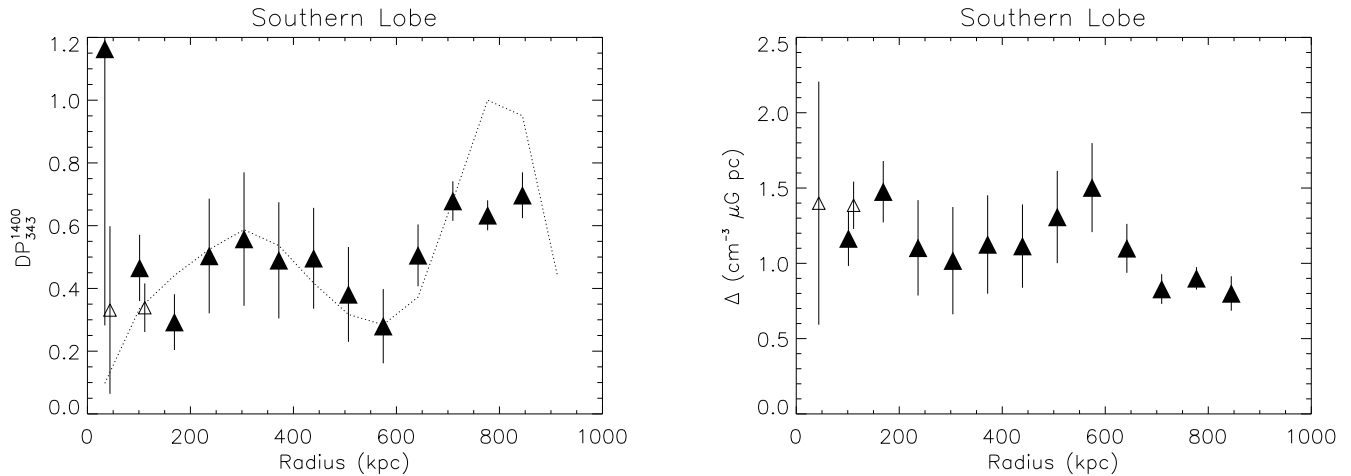

\begin{tabular}{ll}
\psfig{figure=7509.14a,width=0.5\textwidth,angle=90} & \psfig{figure=7509.14b,width=0.5\textwidth,angle=90}\\
\end{tabular}
\caption{Radial profiles of ({\bf a}) the depolarization parameter $DP{_{343}^{1400}}$, and ({\bf b}) the therefrom derived Faraday dispersion $\Delta$ towards the southern radio lobe. The dotted line in the left plot is the radial profile of the total intensity. The open symbols at small radii are the values after the unpolarized core has been subtracted from the data. For clarity, they have been shifted by a small amount in radius.}
\label{fig:faraday_dispersion}
\end{figure*}

To find the origin of the observed depolarization, it is convenient to convert the depolarization parameter $DP$ into a more physical parameter, the so-called {\sl Faraday dispersion}, denoted by $\Delta$. It is defined as the dispersion in the Faraday function $F(\phi) = \int n_e B_{\parallel} dl$, the integral over the line-of-sight of the product of the electron density and the component of the magnetic field along the line-of-sight (Burn 1966).
By lack of knowledge about the true Faraday function, it is usually assumed that $F(\phi)$ has a Gaussian distribution, with standard deviation $\Delta$. It can then be shown that, for a source with a redhift of $z$, the fractional polarization $m$ at observing wavelength $\lambda$, $m(\lambda) = m(0)\exp(-2k^2\Delta^2\lambda^4(1+z)^{-4})$. If $\Delta$ is expressed in units of cm$^{-3}$~$\mu$G~pc, $k=0.81$. In the case of WNB\,0313+683 we thus find $\Delta = 1.33 (-\ln DP{_{343}^{1400}})^{1/2}$~cm$^{-3}$~$\mu$G~pc.\\
Only the southern lobe is well detected in polarized emission at 343 and 1400~MHz. To study the behaviour of the depolarization in the southern lobe as a function of distance from the core, we integrated the total intensity and the polarized flux in annuli of 30\arcsec~width ($\sim$~half a beamsize), centred on the radio core, at both frequencies.  
We excluded the unrelated radio source at RA $03^{\rm h}13^{\rm m}50^{\rm s}$, Dec. $68^{\circ}11'50''$ by blanking it on the maps. Only points where the total intensity in the 343-MHz map exceeded 6~mJy beam$^{-1}$ were used in the integration. For each annulus, the scalar fractional polarization was calculated at each frequency, and from this the depolarization measure $DP_{343}^{1400}$ and the Faraday dispersion $\Delta$.  
A plot of $DP_{343}^{1400}$ as a function of radius from the core is shown in Fig. \ref{fig:faraday_dispersion}. For reference, the dotted line gives the integrated total intensity for each annulus in arbitrary units. Strangely enough, the profile of the depolarization parameter $DP$ follows the line of total intensity quite well.\\  
At large radii ($\ga 700$~kpc, near the southern hotspot) there is only little depolarization. Towards the core there is a small increase, but $DP$ remains more or less constant between 200 and 500 kpc. Close to the core, there is again a decrease in depolarization, but this is due to the influence of the
unpolarized core, as was already explained in Sect. \ref{sec:depol_obs}. The plot of the Faraday dispersion $\Delta$ is also shown in Fig. \ref{fig:faraday_dispersion}. Around the hotspot, $\Delta \approx 0.85$~cm$^{-3}$~$\mu$G~pc. Towards the core it increases to a value of $\approx 1.5$~cm$^{-3}$~$\mu$G~pc at a radius of 600~kpc, and decreases to a value close to $\approx 1$~cm$^{-3}$~$\mu$G~pc at radii between 200 and 500 kpc. Compared to studies of other radio galaxies (e.g. Johnson et al. 1995), the Faraday dispersion towards WNB\,0313+683 is low, even in the innermost 100--200~kpc of the bridges.\\
There are three possible regions where the observed depolarization can occur: First, in the halo of our own galaxy, second, inside the radio source itself, and third, in a large cluster halo surrounding the entire radio source. We will discuss all three possibilities for the case of WNB\,0313+683.\\

\subsection{Depolarization by a galactic foreground screen}

Magneto-ionic clouds in the Galactic plane and halo can depolarize emission from extragalactic radio sources, and cause Faraday effects in the polarization of the synchrotron radiation from our own Milky way. These effects were observed by Wieringa et al. (1993) using the WSRT, and show up in their maps as patches of polarized emission without any related signature in the total power maps. They interpreted these observations by assuming that there must be clouds or filaments of magneto-ionic plasma between the observer and the smooth galactic background emission. Such filaments rotate the polarization angle of the incoming radiation, causing observable structures in the polarized intensity maps of the WSRT interferometer, whereas the total intensity is not changed and, due to its large-scale smoothness, remains undetected by the WSRT.\\
Wieringa et al. (1993) observed that the smallest scale-sizes for these filaments were 5\arcmin--10\arcmin, which is smaller than our radio source. 
However, depolarization is caused by gradients in the Rotation Measure within a single beam. Since the filaments are much larger than our beamsize, we can exclude the galactic halo as the cause of the observed depolarization. Also, if strong gradients were present in the Rotation Measure, they would have shown up in our RM-maps. 
As an extra check we mapped the polarized intensity over the visible area of these observations. We did not detect any large-scale polarized structures, however.\\   

\subsection{Internal depolarization}

Because of the large physical size of WNB\,0313+683, there is considerable pathlength through the source itself. If thermal matter is entrained inside the radio lobes, or if the radio lobes themselves are filamented, internal depolarization is likely to occur.
A classical test for internal depolarization is to observe a breakdown of the linear rotation angle -- $\lambda^2$ relation before a rotation of 90$^{\circ}$ occurs (Laing 1985). In our case, however, most of the rotation is of galactic
origin, and the values of the residual Rotation Measure are not significant (see Sect. \ref{sec:rot_measure_obs}). From a Gaussian fit to the RM distribution over the southern lobe, we found a dispersion in the RM distribution of 0.23~rad~m$^{-2}$. 
For this value, the rotation of the polarization angle with wavelength goes as $\theta - \theta_0 = 13.2\,\lambda^2\,[\degr]$, where $\theta_0$ is the angle at zero wavelength. For a wavelength of 94~cm, which is our lowest frequency channel, the internal rotation of the polarization angle is therefore at most $\sim 12$\degr. This is not high enough for a measurable deviation from the polarization angle -- $\lambda^2$ relation.\\ 
If the depolarization occurs internally, we would also expect a correlation between the amount of observed depolarization and the `depth' of the radio source along the line of sight. This requires knowledge of the source structure along the line of sight, which we do not have, but we can assume cylindrical symmetry around the radio axis.
Since the radio source is `fatter' near the center than near the hotspots, the amount of depolarization towards the center should be larger if it were caused internally. This is under the assumption that there are no large changes in the strength or scale-length of the magnetic field or the density of thermal matter within the source, which can also influence the observed depolarization. Figure \ref{fig:faraday_dispersion} shows that the amount of depolarization indeed increases somewhat at radii below 300~kpc, but this is not highly significant.\\
Internal depolarization is further indicated by a correlation between gradients in the Rotation Measure distribution and the amount of depolarization (Leahy et al. 1986). The RM gradients we find are however small and not significant, so we are not able to test if such a correlation exists in our data.
We therefore have no conclusive evidence supporting internal depolarization, but neither can we exclude it on the basis of the data available.

\subsection{Depolarization by a magneto-ionic halo}

A third mechanism to depolarize radiation is by a large halo surrounding the radio source. A magnetized plasma consisting of thermal ($10^7 - 10^8$~K) electrons and tangled magnetic fields (often called a magneto-ionic plasma) depolarizes radiation passing through it, as long as the scale length of the tangled $B$-field is much smaller than the observing beam (e.g. Laing 1985, Garrington \& Conway 1991). The beamsize we use ($\sim 60''$) translates into a physical size of $\sim 135$~kpc at the distance of WNB\,0313+683. This is much larger than the maximum scale $d_{max}$ of magnetic field tangling in a cluster halo, if this is set by the motion of galaxies through the cluster gas (i.e. 20 kpc; Jaffe 1980). Therefore, we most likely do not resolve the individual `Faraday cells' and do not need to correct for this effect on the observed Faraday dispersion.\\ 
The Faraday dispersion for radiation passing through a depolarizing plasma is given by $\Delta  = \langle n^2\,B_{\parallel}^2\rangle^{1/2}\,(L\,d_{max})^{1/2}$ (e.g. Garrington \& Conway 1991). Here, $L$ is the pathlength through the depolarizing plasma in pc. If a large magneto-ionic halo is responsible for the observed depolarization, we expect an increase in the observed Faraday dispersion at small radii. Typical core radii of the extended X-ray emission in both rich clusters and poor groups of galaxies are 200-600~kpc (Mulchaey \& Zabludoff 1998). 
Radio sources that are at the centers of clusters with strong cooling flows and thus high densities in their central regions have been found to be strongly depolarized (e.g. Taylor \& Perley 1993; Taylor et al. 1994) within their core radius. However, for WNB\,0313+683 there is no strong increase in $\Delta$ at radius $\la 200$~kpc (Fig. \ref{fig:faraday_dispersion}b). Partly, this is caused by the earlier discussed influence of the unpolarized radio core. To remove this effect we carefully subtracted the radio core from the total intensity data. We then recalculated the fractional polarization in the inner annuli. We found that there is still only a small increase in $\Delta$ at distances $\la 200$~kpc, with $\Delta \approx 1.4$~cm$^{-3}~\mu$G~pc. 
Therefore, if there is any cluster gas around the host galaxy, it is either very centrally condensed like a galaxy halo (e.g. Jones \& Forman 1984), or it has a very low density, even at the center of the cluster. Sensitive X-ray observations should be able to disentangle these two possibilities. 
In either case, the heads of the two lobes of WNB\,0313+683 must be currently residing in extremely low-density environments. \\
Earlier we estimated the density of the medium surrounding the radio source using the spectral age of the radio source and assuming ram-pressure at the head of the lobe (Sect. \ref{sec:density_ram}). Using the formula for $\Delta$ given above, we can now try to give an independent estimate. 
Assuming that the radio source resides in a large gaseous halo, the diameter of this halo must be at least comparable to the size of the radio source. We can thus take $L \ga 1$~Mpc, and find $\langle n^2B^2\rangle^{1/2} \la 9.9 \times 10^{-6}$~cm$^{-3}$~$\mu$G.
To estimate the magnetic field strength in the cluster gas, we can use the argument that the pressure of the medium outside the radio source cannot be larger than that in the radio lobes, in order to maintain an equilibrium situation. Using equipartition arguments, this is the same as saying that the magnetic field in the cluster gas cannot be much stronger than the equipartion magnetic field in the radio lobe. From Tab. \ref{tab:energies}, we therefore take a field strength of $0.5~\mu$G as an estimate for the mean magnetic field strength in the cluster gas.
We derive a lower limit for the electron density around the southern lobe, averaged along the line of sight, of $2.0\times 10^{-5}$~cm$^{-3}$. This is an order of magnitude higher than the value derived from ram-pressure arguments (Sect. \ref{sec:density_ram}). However, since both methods depend on many assumptions this is not very surprising.
For instance, if part of the depolarization occurs internally of the source, the density derived above is too high. As we mentioned in the last section, we are not able to exclude this possibility.

\section{The nucleus: a restarted radio source?}

The properties that distinguish WNB\,0313+683 from the general population of GRGs are its inverted spectrum radio core, and the relatively large power in optical emission-lines, as compared to the jet power (see  Sect. \ref{sec:oell}). It is therefore very attractive to believe that these two phenomena are somehow related. We discuss several scenarios to explain these observations. However, a few words of caution are appropriate: The estimates of the jet power are based on many assumptions, such as the validity of equipartition and the correctness of our source age estimate. The power in optical emission-lines, on the other hand, depends strongly on the extinction, which we estimated using the Balmer decrement, but this plays only a minor role in the following discussion.\\ 
Relativistic beaming increases an observed flux (e.g. Blandford \& K\"{o}nigl 1979) and, depending on which component is beamed, can change the observed spectrum. However, for this effect to be important in the core of WNB\,0313+683, the angle of the radio axis to the line of sight must be very small. Also, beamed radio sources, such as BL Lacs, are often highly variable. Further, the optical spectrum is then expected to show a significant contribution of non-thermal (nuclear) continuum, and broad components of the emitted lines would be much more prevalent. 
Our present data are not good enough to exclude core flux variability. However, the continuum in the optical spectrum is dominated by stellar light, and the broad component of the H$\alpha$ and H$\beta$ lines are certainly not as dominant as observed in quasars. This argues against the possibility that the radio jet in the core is oriented close to the line of sight.\\
The inverted spectrum of the core is very suggestive of it being a member of the class of Gigahertz Peaked Spectrum (GPS) radio sources, although the peak in the spectrum is at too high a frequency to be classified as such. The radio structure of GPS sources has been investigated in detail with VLBI (e.g. Snellen 1997), and often show, on parsec scales, a morphology very similar to kpc-sized radio galaxies. 
This has led to the suggestion that they are very young radio sources, just starting to work their way through the ISM of the host galaxy (Mutel \& Phillips 1988). As they grow larger, their peak frequency decreases (Snellen 1997). Therefore, the high peak frequency of the core of WNB\,0313+683 is indicative for a very small radio source which in the view of Mutel \& Phillips must also be very young. But how can such a radio source be the core of a Mpc-sized and very old radio galaxy?\\ 
This might be explained by a halted and restarted jet formation in the nucleus. An estimate of the age of the new source can be provided by the observation that the hotspots in the southern lobe of WNB\,0313+683 are still very luminous. Once the inflow of jet material stops, the hotspots should fade rapidly, on timescales of order $10^4$~yrs (e.g. Reynolds \& Begelman 1997). 
Their presence indicates that they must still be energized. The light travel time from the core to the southern hotspot is almost 3~Myrs, so the hotspot will still receive jet material for at least this amount of time after the central engine has stopped producing the jet.
The new radio source, which we observe as the inverted spectrum core, therefore has a time slot of a few Myrs to form itself. Spectral ages of other radio sources, such as the well known radio source Cygnus~A with an estimated age of only 6~Myrs (Carilli et al. 1991), indicate that it is well possible to produce even large radio sources within a few Myrs.\\ 
Further, the newly formed jet may be much more powerful than the jet that formed the old Mpc-sized radio source. This could explain the discrepancy between the estimated jet power and the emitted power in emission lines: both belong to two different epochs in the lifetime of this source, and have therefore no relation to each other.\\
GPS sources may also be `frustrated' radio sources, embedded in too dense an environment to allow the jet to escape (e.g. O'Dea et al. 1991). In the case of WNB\,0313+683, this would indicate that there must have been a large change in the gas dynamics of the host galaxy, for instance as the result of a merger with a gas-rich galaxy. The optical image we have from the DSS (see Fig. \ref{fig:wsrt1.4+opt}b) is of too poor a quality to investigate the morphology of the host galaxy for any signs of such a phenomenon. A large amount of gas that has been driven into the nuclear region provides fuel for the AGN, so this might explain the strong optical emission lines as well. What the influence on the radio jet would be is however unclear, since there is no good theory yet to investigate this.\\
To investigate a possible `rebirth' of the radio core in more detail, VLBI observations would be extremely interesting. Does the core resemble a small double-lobed radio source? Even more interesting would be a search for superluminal motion in this source, since this would give an important constraint on the orientation of the radio axis. To estimate the amount of gas towards the central engine, a study of HI-absorption would be very valuable.

\section{Summary and conclusions}

We have investigated the newly discovered large radio source WNB\,0313+683. This 15\arcmin~source was discovered in the WENSS. The radio core has been identified with a weak optical galaxy. It has a redshift of $0.0901 \pm 0.0002$, and thus a projected linear size of 2.0~Mpc.
The optical spectrum is typical of a narrow-line radio galaxy. The H$\alpha$-line has a broad ($\sim 6000$~km~s$^{-1}$) component. The color index, determined from the (narrow) H$\alpha$/H$\beta$ line-ratio, is large: $E(B-V) = 0.98 \pm 0.10$~mag, where we have assumed that the extinction is mostly galactic (the galactic latitude is +9.8$^{\circ}$). The HI column density towards the host galaxy, measured in the Leiden-Dwingeloo galactic HI survey, translates into a color index of $0.51$~mag. But the large variations observed on scales of 30\arcmin~in the HI survey maps suggest that denser regions may well exist within a single beam.\\
In addition to the WENSS and the NVSS 1.4-GHz maps, we have observed WNB\,0313+683 with the 100-m Effelsberg telescope at 10.4~GHz, with the WSRT using the 92-cm 8-channel broadband system, and with the VLA at 1.4 and 5 GHz.  
The VLA and Effelsberg observations show that WNB\,0313+683 is a core-dominated radio galaxy. The radio spectral index of the core is $+0.42 \pm 0.03$, at least up to 10.4 GHz, at which frequency it contributes $\sim 25\%$ to the total flux of the source.
The prominence of the core, if due to relativistic beaming, suggests that the radio axis of the (sub)arcsec core structure is oriented at an angle $\la 50\degr$ to the plane of the sky. If the radio axis of the source as a whole is similar to the radio axis of the core, the deprojected linear size would be $\ga$ 2.6 Mpc.\\
Spectral index profiles have been made along the radio axis. They have been used to estimate the age and advance velocities of the two lobes. 
We find that the source has a maximum age of $1.4 \pm 0.1 \times 10^8$ yrs. The found advance velocities of the lobes are $(0.027 \pm 0.002)c$ for the northern lobe, and $\sim (0.03 \pm 0.01)c$ for the southern lobe. The ambient densities have been estimated using ram-pressure equilibrium at the heads of the jet, and by assuming that the impact area is limited to the observed size of the hotspots. We find lower limits to the particle density of $1.6 \times 10^{-6}$~cm$^{-3}$ for the material surrounding the southern hotspot and $4.8 \times 10^{-7}$~cm$^{-3}$ around the northern hotspot.
The contrast in the external density of the two lobes is reflected in the lobe luminosities. The southern lobe is much brighter (by a factor of $\sim\,5$) and appears more confined than the northern one.\\
We have studied the Rotation Measure and depolarization towards WNB\,0313+683. The Rotation Measures have been determined by a new technique, called Rotation Measure Mapping. We find a very smooth RM distribution over the southern lobe, with a mean value of $-10.6 \pm 0.2$~rad~m$^{-2}$. We believe that this is of galactic origin. After subtracting this mean value from the observed values, we only find significant residual RM values towards the central bulge ($-1.9 \pm 0.5$~rad~m$^{-2}$).\\
Depolarization towards WNB\,0313+683 is only important at wavelengths above 21cm, since we measure almost no depolarization between the Effelsberg and the NVSS data but find significant depolarization between the NVSS and the 92-cm WSRT data. 
We have discussed the origin of the depolarization and conclude that it could either be caused inside the radio source itself, or by a large magneto-ionic cluster-like halo surrounding the source. The current data do not allow us to distinguish between these two scenarios. 
We have investigated the distribution of the depolarization, and the derived Faraday depth $\Delta$, towards the southern lobe. We find no strong increase of $\Delta$ near the core, as would be expected if the host galaxy lies at the center of a cluster halo with a dense core, but the values are significantly higher than those near the southern hotspot. Overall, $\Delta$ is close to 1.0~cm$^{-3}~\mu$G~pc.\\ 
In the case that a large halo surrounding the radio source is responsible for the depolarization, we estimate a lower limit for the electron density, averaged along the line of sight towards the southern lobe, of $2 \times 10^{-5}$~cm$^{-3}$. This is an order of magnitude above the values we find in the spectral index analysis. Part of this might be caused by a contribution from internal depolarization. Our data do not exclude this possibility.\\
In many respects, WNB\,0313+683 is like any other GRG. Its size is close to the median value of the `classical' GRGs ($\sim 2$~Mpc), the density of the ambient medium is low, of the order of $10^{-6}$~cm$^{-3}$, the spectral age is large ($\sim 10^8$~yrs), there is only little depolarization, and the Rotation Measures, corrected for the galactic contribution, are low as well ($\la 2$~rad~m$^{-2}$).\\
The inverted spectrum radio core and the very strong optical emission lines as compared to the estimated jet power, suggest an interrupted radio activity of the AGN. VLBI observations are necessary to investigate the structure of the radio core in more detail.\\

\begin{acknowledgements}
KHM was supported by the Deutsche For\-schungs\-ge\-mein\-schaft,
grant KL533/4--2 and by European Commission, TMR Programme, Research Network
Contract ERBFMRXCT96-0034 ``CERES''. 
LL acknowledges Bill Cotton for useful suggestions on the use of NVSS data to study giant radio sources. 
We would like to thank the referee, U. Klein, for many helpful suggestions which considerably improved the paper.     
NRAO is operated by Associated Universities, Inc., under cooperative agreement with the National Science Foundation. 
The INT is operated on the island of La Palma by the Isaac Newton Group in the Spanish Observatorio del Roque de los Muchachos of the Instituto de Astrofisica de Canarias.
The Westerbork Synthesis Radio Telescope (WSRT) is operated by the
Netherlands Foundation for Research in Astronomy (NFRA) with financial
support of the Netherlands Organization for Scientific Research (NWO).
This research has made use of the NASA/IPAC Extragalactic Database (NED) which is operated by the Jet Propulsion Laboratory, California Institute of Technology, under contract with the National Aeronautics and Space Administration. \\

\end{acknowledgements}

\appendix
\section{A new method for finding Rotation Measures}
\label{rot_appendix}

\begin{figure*}
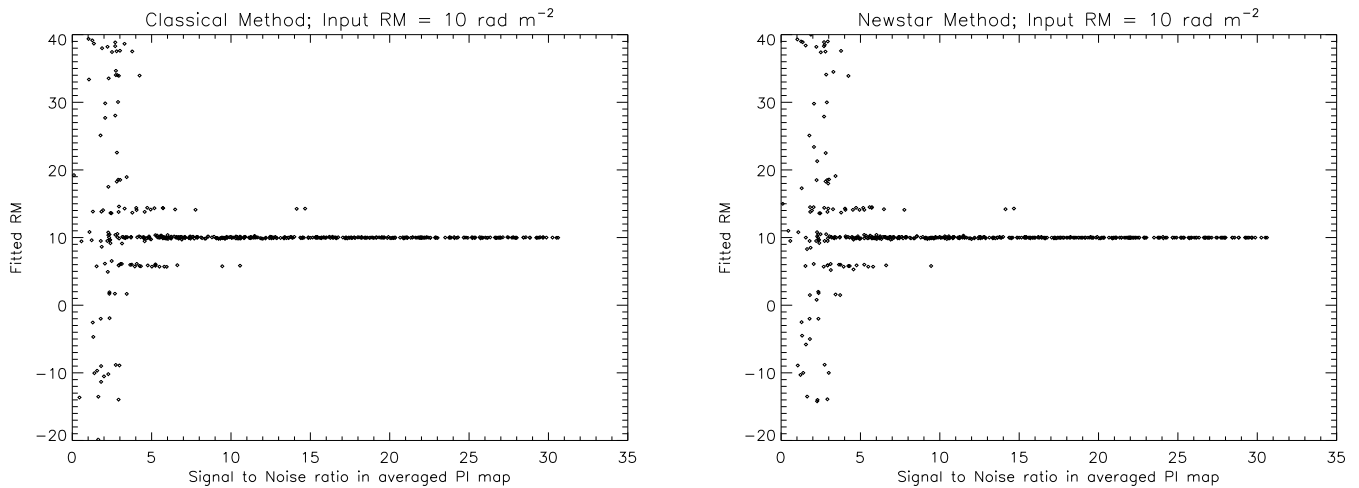

\begin{tabular}{ll}
\psfig{figure=7509.15a,width=0.5\textwidth,angle=90} &\psfig{figure=7509.15b,width=0.5\textwidth,angle=90}\\
\end{tabular}
\caption{\label{fig:rm_simu}Plots showing the result of the simulations of the {\sc newstar} method and the `classical' method for finding Rotation Measures (see text for details). Each plot is a realisation of 500 points with random signal-to-noise values. We used a RM of 10~rad~m$^{-2}$ as input. Both methods give very similar results.}
\end{figure*}

\begin{figure}
\psfig{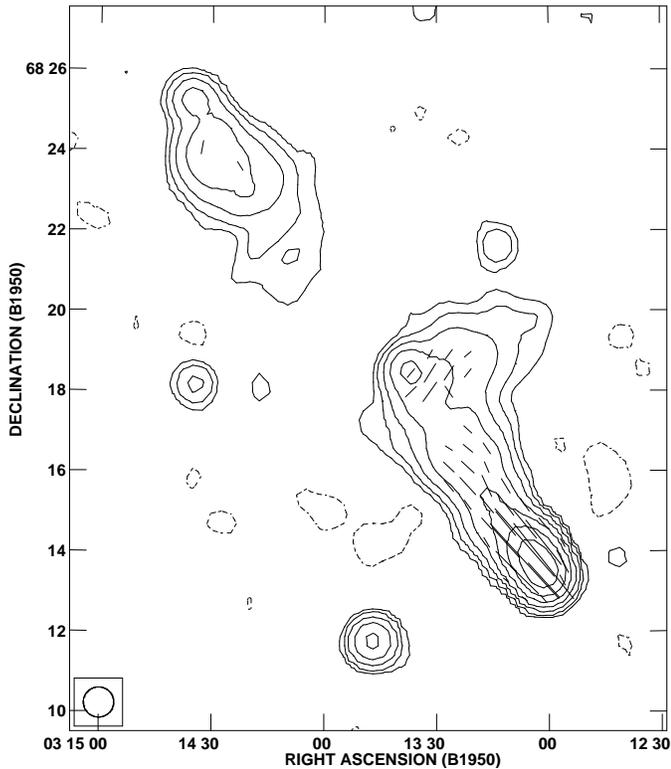}
\caption{\label{fig:nvss_b-field}Contour plot of the WNB\,0313+683 at 1400~MHz from the NVSS survey (see also Fig. \ref{fig:nvss}). The orientation of the vectors represent the orientation of the `true' B-field after rotating the observed $E$-field with the amount of rotation expected from the RM-analysis described in this Appendix. The position angles are almost identical to those inferred from the high-frequency Effelsberg observations. The apparent `wiggle' is caused by the $\sim\!4$~rad~m$^{-2}$ uncertainty that is described in the text.}
\end{figure}

The `classical' method for finding Rotation Measures is to determine the position angle of the polarized emission in each individual channel, and to plot these as a function of the square of the wavelength. The RM is then the slope of the line that fits these datapoints best. This method has proved itself
extensively, but it provides little insight in the spatial distribution of the Rotation Measure when studying extended structures.\\
For this purpose, a new routine has been developed in the NFRA {\sc newstar} reduction and analysis package, called Rotation Measure Mapping. This  routine combines the maps of the Stokes $Q$ and $U$ parameter of each separate frequency channel in the following way. 
Using an `a priori' guess of the RM, the $Q$ and $U$ vectors in each channel are rotated `backwards' over the angle that they would have been rotated over by the assumed RM, relative to the frequency of the highest frequency channel. 
Then, the $Q$ (and $U$) maps of all channels are vector summed. 
Only if the chosen RM is close to the real RM, the rotated $Q$ (and $U$) vectors in each channel will be the same, and the length of the summed vectors will be maximal. Using these vector summed $Q$ and $U$ vectors, which we will refer to as $Q_s$ and $U_s$, respectively, we can calculate the {\em summed} polarized intensity $PI_s = \sqrt{Q^2_{s} + U^2_{s}}$, which will also be at its maximum when the correct RM is chosen. 
Using a range of possible RMs, the RM for which the summed polarized intensity is maximal, must be closest to the real value.\\
The main advantage of the {\sc newstar} method is that maps of summed polarized intensity can be made for a wide range of RMs. These can be displayed in a movie form, so that for extended sources, like WNB\,0313+683, one can look for systematic changes in the RM over the source.
But it can also be used for a more strict analysis of the data.\\
In the classical method, the spectral index and depolarization of the emission, when using widely separated frequencies, will affect the error in the measured polarization angle, and thus the accuracy of the result. In what manner exactly strongly depends on the nature of the source observed.
The intrinsic {\em degree} of polarization for synchrotron radiation with a powerlaw spectrum is not frequency dependent. Therefore, in radio sources with negative spectral slopes, e.g. $S_{\nu} \propto \nu^{-0.8}$, the drop in total intensity at a higher frequency can, at least partly, be counteracted by the lower amount of depolarization at that frequency.\\
In order to test the accuracy and reliability of the {\sc newstar} method, we have performed computer simulations, in which we analysed the same artificial input data set with the {\sc newstar} and the classical method. As input we provided 500 random $Q$ and $U$ intensities, which we assumed to have a frequency of 1400~MHz (cf. the NVSS survey). For each datapoint, we then determined the polarization angle at 1400~MHz and calculated the polarization angles at the other frequencies, using a fixed Rotation Measure of 10~rad~m$^{-2}$. From this, the $Q$ and $U$ intensities at each frequency have been determined. Then noise was added using a Gaussian noise distribution with a standard deviation equal to the rms noise of our real $Q$ and $U$ data.\\     
We find that both methods recover the input RM very well, as long as the polarized intensity $\ga 10\sigma_{PI}$, where $\sigma_{PI}$ is the noise in the averaged polarized intensity map corrected for Ricean bias (see Fig. \ref{fig:rm_simu}). For polarized intensities $\la 8\sigma_{PI}$, the input RM is not reliably recovered. Especially note the two `sidelobes' in the RM distribution, which are separated by $\sim\!4$~rad~m$^{-2}$ from the input RM. They are the result of the $n\pi$ rad ambiguity of the determination of the polarization angle in the high frequency (NVSS) data. Due to the added noise, chosing a value which is 180$^{\circ}$ higher or lower sometimes gives a better fit through the datapoints. We find that $\sim 10\%$ of the datapoints in our simulations which have polarized intensities between 8 and $30\,\sigma_{PI}$ lie in these two sidelobes.\\
We have used this method to correct for the observed position angles of the $E$-field in the 1400-MHz NVSS observations. The result is shown in Fig. \ref{fig:nvss_b-field}. This should be compared with the orientation of the magnetic field that is observed at 10.45~GHz with the Effelsberg telescope (i.e. the vectors plotted in Fig. \ref{fig:effelsberg}a rotated by 90\degr\,). The $\sim\!4$~rad~m$^{-2}$ ambiguity causes the small wiggle in the B-field vectors in the southern lobe of WNB\,0313+683.  

\end{document}